\documentclass[%
 reprint,
 amsmath,amssymb,
 aps,
]{revtex4-2}

\usepackage{graphicx}
\usepackage{dcolumn}
\usepackage{bm}
\usepackage{comment}
\usepackage{physics}
\usepackage{amsmath}
\usepackage{upgreek}
\usepackage{lipsum}

\usepackage{setspace}
\usepackage[dvipsnames]{xcolor}

\begin{document}

\preprint{APS/123-QED}

\title{Quantum control of the tin-vacancy spin qubit in diamond}
\author{Romain Debroux\textsuperscript{1,*}}
\author{Cathryn P. Michaels\textsuperscript{1,*}}
\author{Carola M. Purser\textsuperscript{1,3,*}}
\author{Noel Wan\textsuperscript{2}}
\author{Matthew E. Trusheim\textsuperscript{2,4}}
\author{Jes\'{u}s Arjona Mart\'{i}nez\textsuperscript{1}}
\author{Ryan A. Parker\textsuperscript{1}}
\author{Alexander M. Stramma\textsuperscript{1}}
\author{Kevin C. Chen\textsuperscript{2}}
\author{Lorenzo de Santis\textsuperscript{2,5}}
\author{Evgeny M. Alexeev\textsuperscript{1,3}}
\author{Andrea C. Ferrari\textsuperscript{3}}
\author{Dirk Englund\textsuperscript{2,$\dagger$}}
\author{Dorian A. Gangloff\textsuperscript{1,$\dagger$}}
\author{Mete Atat\"ure\textsuperscript{1,$\dagger$}
 \\ \, \\
}

\affiliation{\textsuperscript{1}Cavendish Laboratory, University of Cambridge, JJ Thomson Ave., Cambridge CB3 0HE, UK}
\affiliation{\textsuperscript{2}Department of Electrical Engineering and Computer Science, Massachusetts Institute of Technology, Cambridge, MA 02139, USA}
\affiliation{\textsuperscript{3}Cambridge Graphene Centre, University of Cambridge, Cambridge CB3 0FA, UK}
\affiliation{\textsuperscript{4}CCDC Army Research Laboratory, Adelphi, MD 20783, USA}
\affiliation{\textsuperscript{5}QuTech, Delft University of Technology, PO Box 5046, 2600 GA Delft, The Netherlands
}

\begin{abstract}
Group-IV color centers in diamond are a promising light-matter interface for quantum networking devices. The negatively charged tin-vacancy center (SnV) is particularly interesting, as its large spin-orbit coupling offers strong protection against phonon dephasing and robust cyclicity of its optical transitions towards spin-photon entanglement schemes. Here, we demonstrate multi-axis coherent control of the SnV spin qubit via an all-optical stimulated Raman drive between the ground and excited states. We use coherent population trapping and optically driven electronic spin resonance to confirm coherent access to the qubit at 1.7\,K, and obtain spin Rabi oscillations at a rate of $\Omega/2\pi=$~3.6(1)\,MHz. All-optical Ramsey interferometry reveals a spin dephasing time of $T_2^*$\,=\,1.3(3)\,$\upmu$s and two-pulse dynamical decoupling already extends the spin coherence time to $T_2$\,=\,0.33(14)\,ms. Combined with transform-limited photons and integration into photonic nanostructures, our results make the SnV a competitive spin-photon building block for quantum networks.    
\end{abstract}

\maketitle

\section{Introduction}

A light-matter quantum interface combines deterministic and coherent generation of single photons with a long-lived matter qubit \cite{Gao2015, Atature2018, Awschalom2018, Wolfowicz2021}. This combination constitutes a foundational building block for quantum networking systems that exploit far-field radiation to generate remote entanglement and near-field interactions to realize nonlinear photonic gates \cite{Kimble2008, Wehner2018, Lindner2009a}. Candidate systems include isolated atoms \cite{Wilk2007,Chou2005,Thompson2013,Schupp2021} and solid-state spins in the optical domain \cite{Senellart2017,Christle2017,Humphreys2018}, as well as superconducting quantum circuits in the microwave regime \cite{Wallraff2020}. An efficient quantum emitter is correspondingly well-suited for light-assisted manipulation of its internal degrees of freedom \cite{Arimondo1976, Meekhof1996, Gupta2001}. The optical domain offers the critical advantage of wireless control fields which can be confined spatially to an optical wavelength, allowing for selective control of individual systems on that length scale, and high-speed control arising from a high electric field density coupling to typically large electric dipole moments \cite{Press2008,Bodey2019,Levine2018,Debnath2016}. 

Diamond stands out as a particularly promising solid-state host for quantum light-matter interfaces \cite{Aharonovich2014DiamondNanophotonics}, enabling all-optical control \cite{Golter2014, Becker2016, Zhou2017, Goldman2020}. Within this material platform, the nitrogen-vacancy center (NV) has been used for pioneering quantum networking tasks owing to its excellent spin coherence \cite{Togan2010, Childress2013, Hensen2015, Bradley2019, Pompili2021}. Scaling up faces the challenge of improving its optical performance with tailored nanostructures \cite{Riedel2017}, which remains difficult owing to the NV's sensitivity to nearby surfaces as a result of its permanent electric dipole moment \cite{Faraon2012, Bernien2012}. In contrast, the group-IV color centers \cite{Bradac2019, Hepp2014, Rogers2014, Rose2017, Zhang2020, Siyushev2017, Rugar2020, Rugar2020a,Trusheim2020, Trusheim2019} are naturally compatible with photonic nanostructures owing to their inversion symmetry \cite{Thiering2018, Bradac2019}, and collection efficiencies exceeding 90\% have been recently demonstrated \cite{Bhaskar2020,Kuruma2021, rugar2021, fuchs2021}. Of these, the negatively charged silicon-vacancy center (SiV) is the most studied, with demonstrations of coherent control of its ground state by microwave \cite{Pingault2017}, all-optical \cite{Becker2018}, and acoustic \cite{Maity2020} drive techniques. At millikelvin temperatures, where dephasing due to single-phonon scattering between orbital levels is suppressed, coherence times up to 13\,ms \cite{Sukachev2017} allow for more mature demonstrations of entanglement \cite{Evans2018, Nguyen2019, Bhaskar2020}. Building on these achievements, the recently reported tin-vacancy center (SnV) \cite{Rugar2020, Rugar2020a, Trusheim2020} shares the desirable optical properties of SiV and provides the additional advantages of (1) a long spin lifetime of $10$\,ms at 3.25\,K (extrapolated to $>$1\,s at 1.7\,K) \cite{Trusheim2020} and (2) optical cyclicity in the presence of an off-axis magnetic field or strain, which can allow for simultaneous single shot readout and efficient coupling to nuclear spins. These advantages stem from a large spin-orbit coupling, which suppresses decoherence due to single phonon scattering in the ground state and establishes a common quantization axis between the ground and excited states, providing robust spin cyclicity. Conversely, the strong spin-orbit coupling also gives rise to orbital-forbidden spin transitions which has limited microwave based spin control \cite{Trusheim2020} and has cast doubt on the feasibility of fast, coherent control of the SnV spin qubit.

\begin{figure*}
\centering
\includegraphics[width=2\columnwidth]{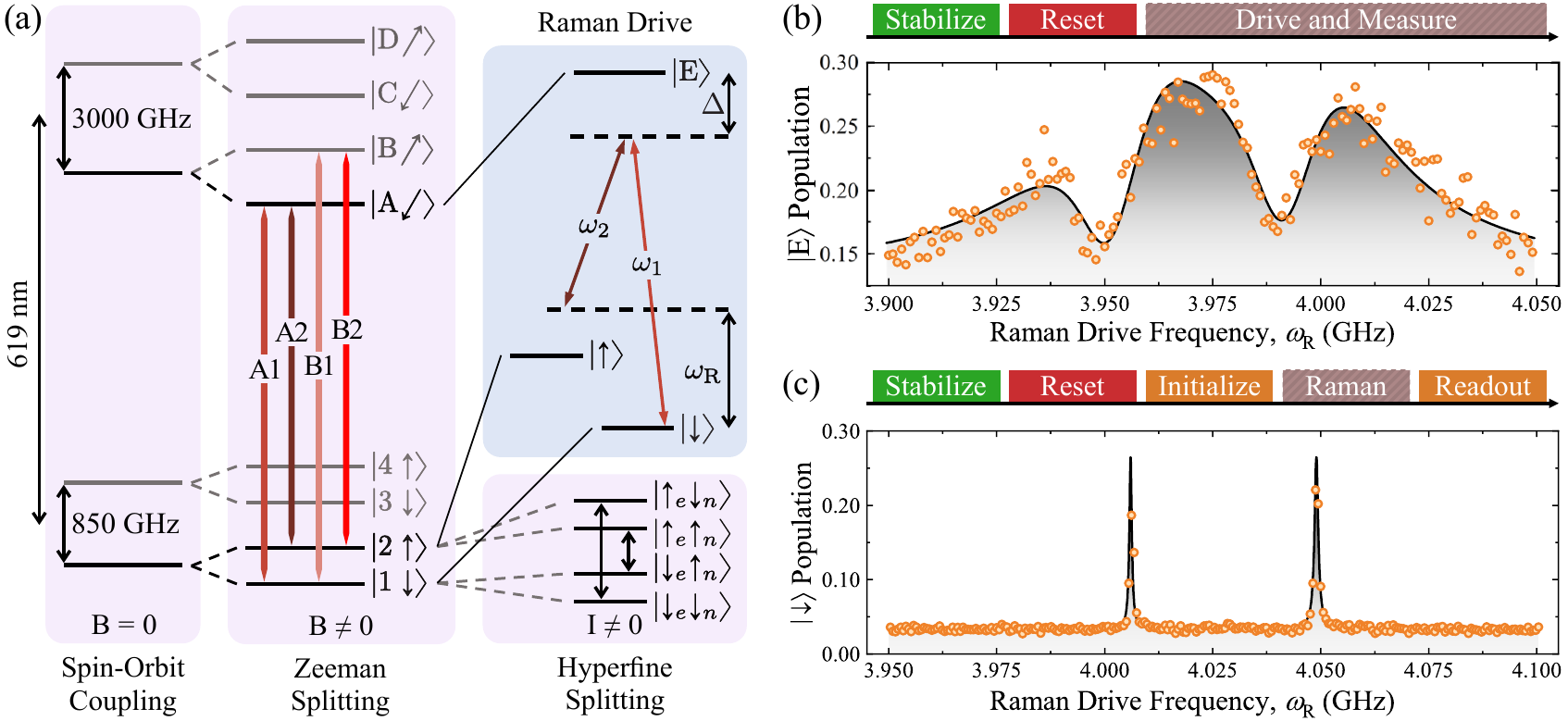}
\caption{\textbf{SnV and Raman drive}. (a) Lilac panels: Energy levels of the SnV split by spin-orbit coupling (left), Zeeman splitting (middle), and hyperfine splitting (right). Blue panel: Qubit defined by the $\ket{\downarrow}$ and $\ket{\uparrow}$ electronic levels. An optical lambda scheme is defined by two optical fields, $\omega_1$ and $\omega_2$ detuned by $\Delta$ from the excited state $\ket{\textrm{E}}$. The Raman drive frequency $\omega_\textrm{R}$\,=\,$\omega_1$\,-\,$\omega_2$ is the energy offset between the two fields. (b) $\ket{\textrm{E}}$ population plotted in orange circles as a function of $\omega_\textrm{R}$ for $\Delta = 0$ and 200\,mT magnetic-field strength applied at 54.7$^\circ$ relative to the SnV symmetry axis. The pulse sequence consists of a $\sim$10\,$\upmu$W green stabilization pulse for 50\,$\upmu$s, B2 reset pulse for 30\,$\upmu$s, and Raman drive with $p=40(4)$\,nW for 2\,$\upmu$s. The solid curve is a 3-level model describing coherent population trapping. (c) $\ket{\downarrow}$ population plotted in orange circles as a function of $\omega_\textrm{R}$ for $\Delta = 600$\,MHz at 204\,mT magnetic-field strength. After the stabilize and reset pulses, the sequence includes an initialization pulse for 30\,$\upmu$s, a Raman drive with $p=40(4)$\,nW for 1\,$\upmu$s, and a readout pulse on A1 for 30\,$\upmu$s. The solid curve is a two-Lorentzian fit resulting in an average linewidth of $900(200)$\,kHz and spin resonances split by $42.6(4)$\,MHz.}
\label{fig1}
\end{figure*}

In this letter, we demonstrate all-optical multi-axis coherent control of the SnV spin qubit by driving its efficient and coherent optical transitions with microwave-modulated laser fields. We demonstrate the flexibility of the all-optical approach by implementing coherent population trapping, optical Rabi driving, Ramsey interferometry and dynamical decoupling of the SnV spin qubit. We further measure an electron-nuclear hyperfine coupling strength of 42.6(4)\,MHz for a spin-active Sn isotope. These results confirm the promise of SnV as a competitive next-generation light-matter quantum interface.

\section{COHERENT OPTICAL ACCESS TO THE SnV SPIN QUBIT}

The lilac frames in Fig.\,1a illustrate the formation of the energy level structure for the negatively charged SnV under the spin-orbit, Zeeman, and hyperfine couplings (SI I). The strong spin-orbit coupling results in the ground- and excited-state manifolds having two orbital levels split by 850\,GHz and 3000\,GHz, respectively \cite{Trusheim2020}, with a 484\,THz (619\,nm) optical transition energy between the manifolds. An external magnetic field lifts the degenerate spin-orbit states via the Zeeman effect. Our qubit is defined as the Zeeman-split electronic spin states in the lower orbital branch, $\ket{1 \downarrow}$ and $\ket{2 \uparrow}$. Spin-orbit interaction sets spin quantization along the SnV crystallographic symmetry axis for both ground and excited states. However, this axis pinning is weaker in the ground state manifold and its spin quantization axis can be perturbed by strain or magnetic field applied perpendicular to the SnV symmetry axis \cite{Hepp2014} with negligible impact on the excited state manifold. This mismatch between the quantization axes of the ground and excited spin states allows for the spin-cycling transitions (A1 and B2) to achieve single-shot optical readout of the qubit, in tandem with the spin-flipping transitions (A2 and B1) to realize an optical lambda scheme (SI I). The branching ratio $\eta$ between the spin-cycling and the spin-flipping relaxation rates can be controlled by the strength of the applied perturbation, and we operate with $\eta$\,$\approx$\,100 in this work (SI II). The right-hand lilac frame displays the effect of hyperfine interaction. With 16.6\% natural abundance, the Sn host atom can be a spin-active isotope ($^{115}$Sn, $^{117}$Sn, or $^{119}$Sn), which couples the electronic spin qubit to the $I$=1/2 nuclear spin.

The light blue frame in Fig.\,1a highlights how we leverage the SnV optical transitions to realize an optical lambda scheme between the $\ket{1 \downarrow}$ and $\ket{2 \uparrow}$ qubit states ($\ket{\downarrow}$ and $\ket{\uparrow}$, respectively) and the excited state $\ket{\textrm{E}}$. This is achieved by simultaneously driving the A1 and A2 transitions with lasers at frequencies $\omega_1$ and $\omega_2$, respectively, detuned relative to one another by the Raman frequency $\omega_\textrm{R} = \omega_1-\omega_2$. The Raman scheme is further detuned from the excited state by the single-photon detuning $\Delta$. For unpolarized light, the Rabi rate at which the spin is driven is then $\Omega = \frac{1}{\sqrt{\eta}}\frac{p}{p_\textrm{sat}}\frac{\Gamma^2}{4\Delta}$, where $p$ is the power in each of the optical fields driving A1 and A2, $p_\textrm{sat}\approx5$\,nW is our emitter's saturation power for the spin-cycling transition, and $\Gamma / 2\pi$\,=\,35\,MHz is the excited state relaxation rate\cite{Foot} (SI III). Typical experimental values of order $p/p_\textrm{sat}$\,=\,10 and $\Delta/2\pi$\,=\,300\,MHz, and a measured $\eta$\,=\,80(5) at 0.2-T magnetic field (SI II), place the spin Rabi frequency in the MHz-scale, which comfortably exceeds the inhomogeneous dephasing rate $1/T_2^*\approx1$\,$\upmu$s$^{-1}$ \cite{Trusheim2020}, as required for coherent spin control. Driving the spin optically also causes a detuning-dependent excited-state scattering rate $\Gamma_\textrm{os}$\,=\,$\frac{p}{p_\textrm{sat}} \frac{\Gamma^3}{8\Delta^2}$ \cite{Foot}, which introduces a spin relaxation rate $1/T_{1,\textrm{os}}$\,=\,$\Gamma_\textrm{os}/\eta$ and a spin dephasing rate $1/T_{2,\textrm{os}}$\,=\,$\Gamma_\textrm{os}$ (SI III). Maximizing the fidelity of a $\pi/2$ gate with respect to $\Delta$ we find an optimal balance when $\Gamma_\textrm{os} = 1/T_2^*$ (SI III).

Our SnV device consists of a nanopillar array fabricated into an Sn$^+$ ion-implanted diamond (SI IV) and is cooled to 1.7\,K in a magneto-optical cryostat (SI V) \cite{Trusheim2020}. Our all-optical measurement sequences include stabilization, reset, initialization, Raman drive and readout pulses. The \textit{stabilize} pulse uses a 532-\,nm laser to stabilize the charge environment of the SnV. The \textit{initialize} (\textit{reset}) pulse consists of resonantly driving the A1 (B2) transition, which polarizes the SnV spin into $\ket{\uparrow}$ ($\ket{\downarrow}$) state in a time $\eta/\Gamma$ ($\approx$1\,$\upmu$s) via relaxation through the weakly allowed spin-flipping transition A2 (B1), achieving up to 99\% initialization fidelities (SI II). The reset pulse polarizes population into the $\ket{\downarrow}$ state immediately before the initialization pulse, such that the fluorescence intensity of the initialization pulse corresponds to $\sim$100\% of the population. The fluorescence intensity of the \textit{readout} pulse, resonant on A1 and normalized to that of the initialization pulse then provides a direct measurement of the population in the $\ket{\downarrow}$ state. The \textit{Raman drive} encompasses all coherent pulse combinations we use in this work relying on a stimulated Raman process. For the optical pulses within the Raman drive, $\omega_{1,2}$ are realized using the two sidebands generated by passing a single-frequency laser through a microwave (MW)-modulated electro-optic modulator. The MW modulation frequency splits the sidebands by $\omega_{\textrm R}$, the MW modulation amplitude determines the power $p$ in each of the $\omega_{1,2}$ sidebands, and the MW modulation phase dictates the relative phase between $\omega_{1,2}$ sidebands, the phase of the Raman drive $\phi$.

Figure 1b displays a coherent population trapping (CPT) measurement as a first step to verify coherent optical access to the SnV spin qubit, where $\omega_1$ and $\omega_2$ sidebands drive the spin-conserving and spin-flipping transitions, respectively, for $\Delta$\,=\,0. The top panel shows the CPT pulse sequence, in which initialization, drive, and readout pulses are combined into one \textit{drive and measure} step. The main panel presents the steady-state SnV fluorescence during the drive pulse as a function of $\omega_{\textrm R}$. We observe a broad feature, whose width is comparable to the excited state linewidth $\Gamma$, accompanied by two narrower dips. A narrow dip in the SnV fluorescence spectrum (SI VI) corresponds to the generation of a dark coherent superposition of the two ground states, $\frac{1}{\sqrt{1+\eta}}[ \ket{\downarrow} - \sqrt{\eta} \ket{\uparrow}]$ (SI VI), and is obtained when $\omega_{\textrm R}$ matches the spin-splitting frequency. In the presence of a spin-active Sn isotope ($I$\,=\,1/2), the CPT resonance splits into two dips arising from two nuclear-spin preserving transitions that are separated by the corresponding hyperfine coupling rate. The CPT spectrum in Fig.\,1b indicates that the electronic spin qubit of the single SnV color center we measure, confirmed via intensity-correlation measurements (SI VII), is indeed coupled to a spin-1/2 nuclear spin. Fitting a theoretical model (black curve) to the CPT data, using a Lindbladian master-equation formalism (SI VI), reveals a hyperfine coupling strength $\sim$\,$40$\,MHz, commensurate with previous reports on other group-IV color centers \cite{Pingault2017, Rogers2014}. The depth of the two CPT resonances confirms that the coherences of the spin ground states and the optical transitions are sufficient to implement coherent optical drive \cite{Rogers2014}.

Having identified the two spin resonances via CPT, we move to the far-detuned stimulated Raman regime $\Delta$\,$\gg$\,$\Gamma$ to suppress scattering from the excited state during the coherent drive sequence. The top panel of Fig.\,1c shows the pulse sequence, where the initialization, drive, and readout pulses are now separate operations. The main panel of Fig.\,1c presents the population recovery of the $\ket{\downarrow}$ state as a function of $\omega_{\textrm R}$. When $\omega_{\textrm R}$ matches one of the electronic-spin resonances, population from the initial $\ket{\uparrow}$ state is transferred to the $\ket{\downarrow}$ state, resulting in a peak in the $\ket{\downarrow}$ population. We fit our hyperfine-split double-peaked spectrum with two independent Lorentzian lineshapes (solid curve), thereby allowing for high-precision measurement of the hyperfine constant for this Sn isotope, $A$\,=\,42.6(4)\,MHz. The all-optical excitation of a single electronic-spin transition $\omega_e$ (defined as the lower energy peak in Fig.\,1c) sets the stage for its coherent control. 

\section{MULTI-AXIS COHERENT CONTROL}

\begin{figure}
\centering
\includegraphics[width=1\columnwidth]{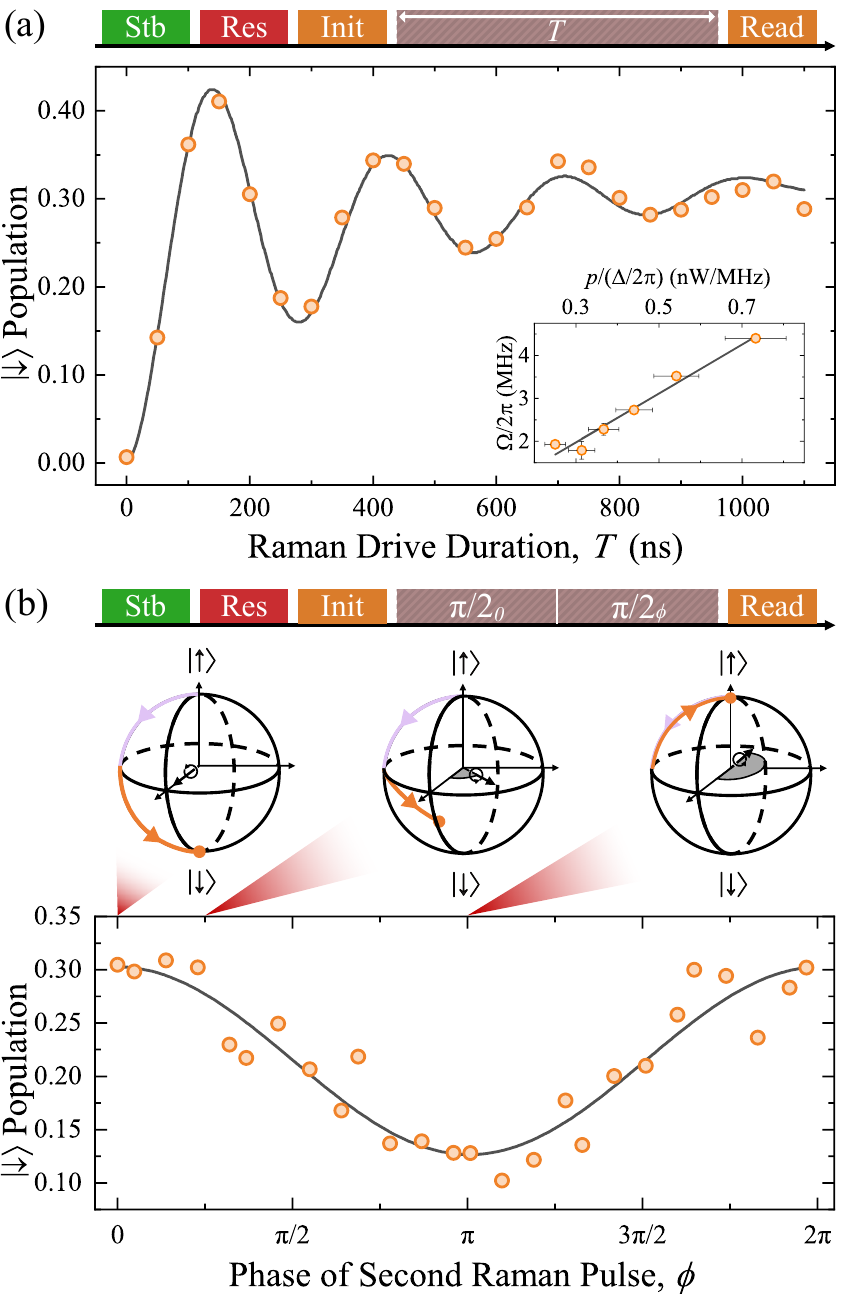}
\caption{\textbf{Multi-axis coherent spin qubit control}. (a) $\ket{\downarrow}$ population (orange circles) as a function of the Raman drive duration $T$ with the pulse sequence shown at the top. The Raman drive is applied with $\Delta/2\pi$\,=\,1.2\,GHz and $p$\,$=$\,650(70)\,nW. The black curve is a fit to a 2-level model under a master equation formalism (SI III). Inset: $\Omega$ as a function of $p/\Delta$ with a linear fit to the data (solid curve). (b) Pulse sequence (top) with one $\pi/2$ pulse about $x$ and a second about an axis rotated by an azimuthal angle $\phi$ from the $x$-axis. The $\pi/2$ pulse duration is determined from Rabi measurements taken with $\Delta/2\pi$\,=\,300\,MHz and $p$\,=\,260(30)\,nW. Illustrated on the Bloch spheres are trajectories for $\phi$\,=\,0 (left), $\phi$\,=\,$\pi/4$  (center), and $\phi$\,=\,$\pi$ (right). The $\ket{\downarrow}$ population (orange circles) is plotted as a function of $\phi$. The solid curve is a cosine function.}
\end{figure}

We now demonstrate coherent spin control in the stimulated Raman regime. Figure\,2a shows the Rabi oscillations of the population in the $\ket{\downarrow}$ state as we sweep the drive pulse duration $T$, with $\omega_\textrm{R}$\,=\,$\omega_e$. By fitting the data to a two-level model under a master equation formalism (solid curve), we extract a Rabi rate of $\Omega / 2\pi$\,=\,3.6(1)\,MHz (SI III). This is an improvement of nearly three orders of magnitude in spin Rabi frequency over direct microwave control realized thus far for SnVs \cite{Trusheim2020} (SI I). The inset shows the dependence of Rabi rate on power and detuning, with the expected linear dependence $\Omega$\,$\propto$\,$p/\Delta$. Our model further yields a spin dephasing rate $\Gamma_\textrm{os}$\,=\,7(4)\,$\upmu \textrm{s}^{-1}$, in good agreement with the expected scattering rate at a detuning of $\Delta/2\pi$\,=\,1200\,MHz. Taken together, $\Omega$ and $\Gamma_\textrm{os}$ directly translate to a $\pi/2$-gate fidelity of 92(4)\%. The same mechanism sets an upper limit on the fidelity of all subsequent measurements involving more complex pulses. While the pulse fidelity achieved here remains modest, operating at larger detuning with increased laser power places high-fidelity gates within reach (SI III).

Full qubit control requires coherent drive about an arbitrary axis. Our approach realizes this via the combination of the two-photon detuning, $\delta$\,=\,$\omega_\textrm{R} -\omega_e$, and $\phi$, the MW-controlled phase between the $\omega_{1,2}$ sidebands. The latter sets the control axis within the equatorial plane of the Bloch sphere, which is particularly relevant for implementing control sequences from the nuclear magnetic resonance toolbox \cite{Schweiger2001, Abragam}. We demonstrate this multi-axis control in Fig.\,2b via a drive sequence comprising two $\pi/2$ pulses, with the first driving the spin about the $x$-axis of the Bloch sphere ($\phi=0$) and the second driving the spin about an axis rotated by an angle $\phi$ away from the $x$-axis. The population of the $\ket{\downarrow}$ state depends periodically on $\phi$ over the full $2\pi$ range, where the cumulative drive for the maximum (minimum) $\ket{\downarrow}$ population corresponds to an effective $\pi$ (0) pulse. The phase dependence of the population readout confirms our ability to choose the quantum state rotation axis.
\section{MEASURING SNV SPIN COHERENCE}

\begin{figure}
\centering
\includegraphics[width=1\columnwidth]{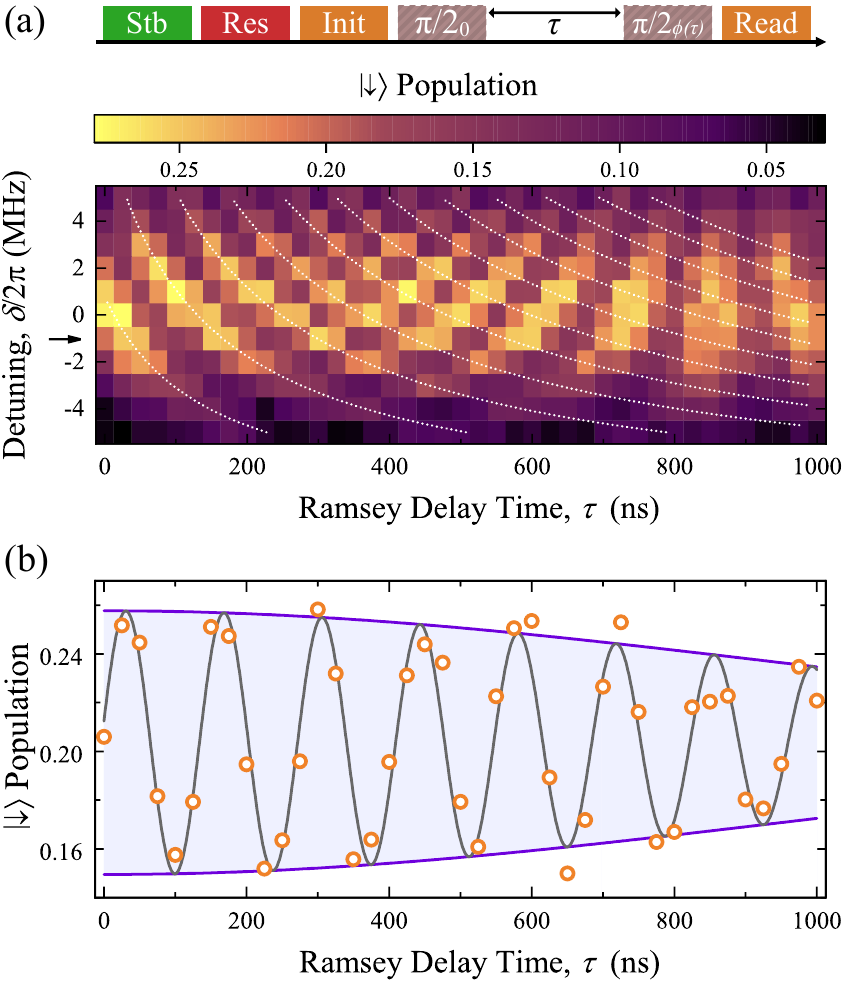}
\caption{\textbf{Ramsey interferometry}. (a) Ramsey pulse sequence with two $\pi/2$ pulses separated by a delay time $\tau$ (top). The phase of the second pulse is swept according to $\phi = \tau\omega_\textrm{S}$, where $\omega_\textrm{S} / 2\pi$\,= 5\,MHz. The color indicates $\ket{\downarrow}$ population plotted as a function of $\tau$ and the two-photon detuning, $\delta$. Dotted white curves provide guide to the eye for the expected $\ket{\downarrow}$ population recovery. Fits of the form $a\exp\left(-(\tau/T_2^*)^2\right)\sin(\omega_{\textrm{Ramsey}}\tau+\alpha)$ at each $\delta$ yield an average $T_2^*$\,=\,1.3(3)\,$\upmu$s and $\Delta_{\textrm{AC}} / 2\pi$\,=\,3.3(5)\,MHz (SI VIII). (b) Linecut at $\delta/2\pi$\,=\,-1\,MHz indicated by an arrow in (a), where the $\ket{\downarrow}$ population (orange circles) is measured as a function of $\tau$. The black curve corresponds to the fit described in (a), where $\omega_{\textrm{Ramsey}}/2\pi$\,=\,7.27(3)\,MHz, $T_2^*$\,=\,1.4(3)\,$\upmu$s, $a$\,=\,0.054(4) and $\alpha$\,=0.16(9). The purple shaded region is bounded by the envelope $\pm a\exp\left(-(\tau/T_2^*)^2\right)$.
}
\end{figure}

We use multi-axis coherent control to implement Ramsey interferometry in order to measure the inhomogeneous dephasing time $T_2^*$ of the SnV spin qubit. The top panel of Fig.\,3a shows the corresponding pulse sequence comprising of two $\pi$/2 pulses separated by a time delay $\tau$. We further impose a periodic recovery of the Ramsey signal by varying the rotation angle $\phi$ for the second $\pi/2$ pulse as a function of $\tau$, such that $\phi = \tau \omega_\textrm{S}$ with the serrodyne frequency $\omega_{\textrm S} / 2\pi$\,=\,5\,MHz. The main panel of Fig.\,3a presents the dependence on $\tau$ and $\delta$ of the $\ket{\downarrow}$ population, which oscillates as a function of $\tau$ with a sum frequency given by $\omega_{\textrm{Ramsey}} = \omega_\textrm{S} + \delta + \Delta_{\textrm{AC}}$, where $\Delta_{\textrm{AC}}$ is the differential AC Stark shift. The latter originates from the $\ket{\downarrow}$ state's stronger coupling to the Raman fields, and is only present during the Raman drive, thus acting as an effective detuning between the free precession rate of the spin and that of the drive's rotating frame (SI VIII). The period of the Ramsey fringes follows the expected $2\pi/\omega_\textrm{Ramsey}$ behavior.

Figure\,3b is an example line cut of $\ket{\downarrow}$ population as a function of $\tau$ for a fixed $\delta / 2\pi$\,=\,-1\,MHz. Fitting with the function $e^{-(\tau/T_2^*)^2}\sin(\omega_{\textrm{Ramsey}}\tau)$ yields $\omega_\textrm{Ramsey}/2\pi$\,=\,7.27(3)\,MHz, and hence $\Delta_{\textrm{AC}}/2\pi$\,=\,3.3(5)\,MHz, comparable to the expected value (SI VIII). We note that gate fidelity reduces the contrast in these measurements, but does not affect the coherent spin precession between the two $\pi/2$ pulses. The Gaussian envelope $e^{-(\tau/T_2^*)^2}$ provides an estimate of the spin inhomogeneous dephasing time $T_2^*$. By applying our model to the data for each $\delta$ in panel a, we extract $T_2^*$\,=\,1.3(3)\,$\upmu$s. This is well within the range of expected inhomogeneous dephasing times limited by the naturally abundant $^{13}$C nuclear spins in diamond \cite{Sukachev2017, Trusheim2020}, and indicates that the SnV coherence is not phonon-limited at 1.7\,K.

\section{IMPLEMENTING DYNAMICAL DECOUPLING}

\begin{figure}
\centering
\includegraphics[width=1\columnwidth]{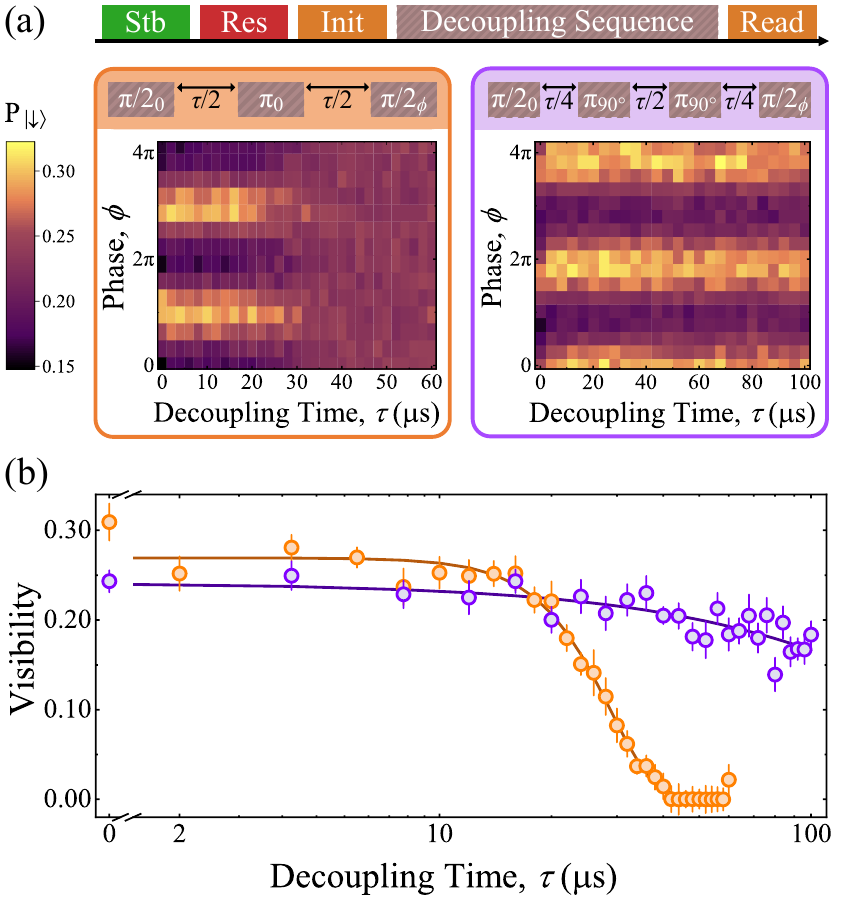}
\caption{\textbf{Dynamical decoupling}. (a) Decoupling pulse sequence (top) with two implementations: (left orange panel) Hahn echo, (right purple panel) CPMG-2. The phase $\phi$ of the second $\pi/2$ pulse is variable. In both panels, color indicates $\ket{\downarrow}$ population, shown as a function of total decoupling time $\tau$ and phase $\phi$. (b) Visibility $a/b$, obtained from fitting the function $a\cos(\phi)+b$ to the data shown in (a) at each delay time $\tau$, plotted as a function of $\tau$. Hahn echo data (orange circles) are fitted to the function $v_0 \exp\left(-(\tau/T_2)^n\right)+v_\infty$, where $v_0$\,=\,0.26(1), $v_\infty$\,=\,0.013(5), $n$\,=\,3.7(4) and $T_2$\,=\,28.3(6)\,$\upmu$s (solid orange curve). CPMG-2 data (purple circles) are fitted to the same function, with $v_0$\,=\, 0.25(1), $v_\infty$\,=\,0.011(4), $n$\,=\,0.8(3), $T_2$\,=\,0.33(14)\,ms (solid purple curve).}
\end{figure}

To prolong the SnV spin qubit coherence beyond the timescale set by the low-frequency magnetic noise of $^{13}$C nuclei, we embed dynamical decoupling protocols within our optical pulse sequence, as illustrated in Fig.\,4a. We implement two example protocols: Hahn echo \cite{Hahn1950} comprising a single rephasing $\pi$ pulse about the $x$-axis (orange frame) and CPMG-2, a Carr-Purcell-Meiboom-Gill sequence \cite{Carr1954} comprising two rephasing $\pi$ pulses about the $y$-axis (purple frame). Sweeping the phase of the final $\pi$/2 pulse $\phi$ from 0 to 4$\pi$ and the decoupling delay time $\tau$ produces the two-dimensional maps of $\ket{\downarrow}$ population in Fig.\,4a. The phase-dependent modulation of the Hahn echo signal lasts for $\sim$30\,$\upmu$s, while the CPMG-2 signal extends significantly longer. Figure 4b presents the extracted visibility for the $\phi$-dependent modulations for both Hahn echo and CPMG-2 protocols as a function of decoupling delay time $\tau$. 
Fitting the Hahn echo visibility (orange circles) as a function of $\tau$ with a stretched exponential function $\exp\left(-(\tau/T_2)^n\right)$ reveals an extended coherence time $T_2$\,=\,28.3(6)\,$\upmu$s. The exponent $n$\,=\,3.6(3) is consistent with a noise spectrum from a slowly evolving nuclear-spin bath in diamond \cite{Childress2006,Sukachev2017}. Applying the same fit function to the CPMG-2 visibility (purple circles) we find an improved coherence time of $T_2$\,=\,0.33(14)\,ms and $n$\,=\,0.8(3). This value for the exponent $n$\,$\approx$\,1 implies that the nuclear-spin bath is no longer the primary dephasing mechanism at this timescale and that an irreversible mechanism dominates. With phonon-induced dephasing expected to be on the order of tens of milliseconds, a more likely source for dephasing is scattering off the excited state during our pulse sequence. Indeed, we find that imperfect laser suppression during the decoupling delay time causes on average one optical scattering event off the excited state within $T_{2,\textrm{os}} \approx 0.2$\,ms (SI IX), which is consistent with our measured CPMG-2 $T_2$. Despite this technical limitation, our CPMG-2 coherence time is already within a factor of 2 of the best reported CPMG-2 coherence time for SiV at 100\,mK \cite{Sukachev2017} and can be prolonged further with straightforward improvements of our instrumentation. 

\section{CONCLUSIONS AND OUTLOOK}
Our all-optical multi-axis coherent control of the SnV spin qubit establishes this alternative diamond color center as an attractive spin-photon quantum interface in the quest for efficient quantum networks. Carrying over the operational advantages that are common to the previously investigated group-IV color centers, such as large Debye-Waller factor \cite{Bradac2019}, transform-limited photon generation \cite{Trusheim2020}, and integration into photonic nanostructures enabled by their symmetry \cite{Wan2020, Nguyen2019a}, the SnV brings two additional advantages. First, the SnV spin remains competitive with the NV benchmark without requiring millikelvin operation temperature. Second, the strength of the spin-orbit interaction in the SnV offers the opportunity to simultaneously perform coherent spin control, single-shot readout and nuclear spin access all via the optical transitions. Our SnV spin coherence time and Rabi rate can both be improved with stronger optical fields, which facilitates the suppression of optical scattering. Gate fidelities can be improved with technical refinements, and tailored pulse protocols are expected to result in 99.6\% gate fidelity for $\pi$-rotations \cite{Takou2021}. An immediate next step towards realizing an efficient quantum memory is extending our all-optical approach to control the intrinsic Sn nuclear spin \cite{Michaels2021}. Further, integrating the SnV into photonic nanostructures \cite{Wan2020, Rugar2020} will increase the photon collection efficiency, and in parallel can strengthen the optical Rabi drive. Such structures should therefore enable efficient coherent control of an electronic spin coupled to a nuclear quantum memory with single-shot readout, a key building block for quantum networks \cite{Kimble2008, Wehner2018}.

\section*{ACKNOWLEDGMENTS}

We acknowledge support from the ERC Advanced Grant PEDESTAL (884745), the EU Quantum Flagship 2D-SIPC, ERC Grants Hetero2D, GSYNCOR, and EPSRC Grants EP/K01711X/1, EP/K017144/1, EP/N010345, EP/L016087/1. R.D. acknowledges support from the Gates Cambridge Trust. J.A.M. acknowledges support from the Winton Programme. R.A.P. acknowledges support from the General Sir John Monash Foundation. A.M.S. acknowledges support from EPSRC/NQIT. L.D. acknowledges funding from the European Union's Horizon 2020 research and innovation program under the Marie Sklodowska-Curie grant agreement No 840393. M.T. acknowledges support through the Army Research Laboratory ENIAC Distinguished Postdoctoral Fellowship. K.C.C. acknowledges funding support by the National Science Foundation Graduate Research Fellowships Program (GRFP) and the NSF STC Center for Integrated Quantum Materials (CIQM), NSF Grant No. DMR-1231319 and NSF Award No. 1839155. N.W. acknowledges funding from the NSF Center for Ultracold Atoms and the NSF Center for Quantum Networks. D.A.G. acknowledges support from a St John's College Title A Fellowship. D.E. acknowledges further support by the MITRE Quantum Moonshot Program. 

\vspace{5pt}
~ \\ 
\textsuperscript{*}\,These authors contributed equally to this work.
\\
\textsuperscript{$\dagger$}\,Correspondence should be addressed to: \\englund@mit.edu, dag50@cam.ac.uk, ma424@cam.ac.uk.


%

\end{document}


\preprint{APS/123-QED}

\title{Supplemental information for \\ 
Quantum control of the tin-vacancy spin qubit in diamond}
\author{Romain Debroux\textsuperscript{1,*}}
\author{Cathryn P. Michaels\textsuperscript{1,*}}
\author{Carola M. Purser\textsuperscript{1,3,*}}
\author{Noel Wan\textsuperscript{2}}
\author{Matthew E. Trusheim\textsuperscript{2}}
\author{Jes\'{u}s Arjona Mart\'{i}nez\textsuperscript{1}}
\author{Ryan A. Parker\textsuperscript{1}}
\author{Alexander M. Stramma\textsuperscript{1}}
\author{Kevin C. Chen\textsuperscript{2}}
\author{Lorenzo de Santis\textsuperscript{2}}
\author{Evgeny M. Alexeev\textsuperscript{3,1}}
\author{Andrea C. Ferrari\textsuperscript{3}}
\author{Dirk Englund\textsuperscript{2,$\dagger$}}
\author{Dorian A. Gangloff\textsuperscript{1,$\dagger$}}
\author{Mete Atat\"ure\textsuperscript{1,$\dagger$}}

\affiliation{\textsuperscript{1}Cavendish Laboratory, University of Cambridge, JJ Thomson Avenue, Cambridge CB3 0HE, United Kingdom}
\affiliation{\textsuperscript{2}Department of Electrical Engineering and Computer Science, Massachusetts Institute of Technology, Cambridge, MA 02139, USA}
\affiliation{\textsuperscript{3}Cambridge Graphene Centre, University of Cambridge, Cambridge CB3 0FA, United Kingdom
\\ \ \\
\textsuperscript{*}\,These authors contributed equally to this work. \\
\textsuperscript{$\dagger$}\,Correspondence should be addressed to: englund@mit.edu, dag50@cam.ac.uk, and ma424@cam.ac.uk. \\}

\maketitle
\tableofcontents

\section{Tin-vacancy electronic structure model}
\label{SI:model}
\subsection{Tin-vacancy Hamiltonian}
The ground and excited states of the tin-vacancy (SnV) are composed of two orbital branches, split by the Jahn-Teller, spin-orbit and strain effects, with two spin sublevels each, split by the Zeeman effect \cite{Hepp2014, Trusheim2020}. In this subsection, we will present the Hamiltonian for the ground state manifold and use it to gain insight about the character of our spin qubit. 

The spin-orbit effect term of the Hamiltonian, expressed with $\{\ket{e_x},\ket{e_y}\}$ spanning the orbital subspace and $\{\ket{\uparrow},\ket{\downarrow}\}$ the spin subspace, is given by \cite{Hepp2014}:
\begin{equation*}
    \begin{aligned}
    H_\textrm{SO} = 
\frac{-\lambda_\textrm{SO}}{2}
\begin{bmatrix} 
0 & i\\
-i & 0
\end{bmatrix}_{\ket{e_x}, \ket{e_y}}
\otimes
\begin{bmatrix} 
1 & 0\\
0 & -1
\end{bmatrix}_{\ket{\uparrow}, \ket{\downarrow}}.
    \end{aligned}
\end{equation*}

The eigenstates resulting from the spin-orbit Hamiltonian are:
\begin{equation*}
\begin{aligned}
&\ket{4}_\textrm{SO} = \frac{1}{\sqrt{2}} \left(\ket{e_x\uparrow}+i\ket{e_y\uparrow}\right) = \ket{e_+ \uparrow}\textrm{, with energy } \lambda_\textrm{SO}/2;\\
&\ket{3}_\textrm{SO} = \frac{1}{\sqrt{2}} \left(\ket{e_x\downarrow}-i\ket{e_y\downarrow}\right) = \ket{e_- \downarrow}\textrm{, with energy } \lambda_\textrm{SO}/2;\\
&\ket{2}_\textrm{SO} = \frac{1}{\sqrt{2}} \left(\ket{e_x\uparrow}-i\ket{e_y\uparrow}\right) = \ket{e_- \uparrow}\textrm{, with energy } -\lambda_\textrm{SO}/2;\\
&\ket{1}_\textrm{SO} = \frac{1}{\sqrt{2}} \left(\ket{e_x\downarrow}+i\ket{e_y\downarrow}\right) = \ket{e_+ \downarrow}\textrm{, with energy }-\lambda_\textrm{SO}/2,
\end{aligned}
\end{equation*}
where $e_\pm = \frac{1}{\sqrt{2}} (e_x \pm ie_y)$. This shows that when only spin-orbit is considered, the two lowest energy eigenstates $\ket{1}$ and $\ket{2}$, which form our qubit, are orthogonal in both the orbital and spin subspaces. Since the spin-orbit effect is the dominant effect in the ground state Hamiltonian with $\lambda_\textrm{SO}=850$\,GHz, we will henceforth express all Hamiltonians in the basis set by the spin-orbit eigenstates $\{\ket{e_+}, \ket{e_-}\} \otimes \{\ket{\uparrow}, \ket{\downarrow\}}$: 
\begin{equation*}
H_\textrm{SO} = 
\frac{\lambda_\textrm{SO}}{2}    \begin{bmatrix} 
1 & 0 & 0 & 0\\
0 & -1 & 0 & 0\\
0 & 0 & -1 & 0\\
0 & 0 & 0 & 1\\
\end{bmatrix}\\.
\end{equation*}

The Zeeman effect acts on both the spin and orbital subspaces. The orbital Zeeman effect is heavily quenched by the Jahn-Teller interaction \cite{Hepp2014} and therefore is commonly neglected \cite{Trusheim2020,Maity2020}, as it will be in this work. The spin Zeeman effect is split into the contribution due to magnetic field parallel ($B_z$) and perpendicular ($B_x, B_y$) to the SnV spin-orbit axis. The Hamiltonian for the parallel field is given by: 
\begin{equation*}
    \begin{aligned}
    H_{Z, \parallel} = 
\frac{\gamma_e}{2}
\begin{bmatrix} 
1 & 0\\
0 & 1
\end{bmatrix}
\otimes
\begin{bmatrix} 
B_z & 0\\
0 & -B_z
\end{bmatrix}
    \end{aligned}
\end{equation*}
with $\gamma_e = 2\mu_B/\hbar$ denoting the electron gyromagnetic ratio. This Hamiltonian is diagonal in the spin-orbit basis, and therefore leaves the eigenstates unchanged. The Hamiltonian for the perpendicular magnetic field is given by: 
\begin{equation*}
    H_{Z, \perp} = 
\frac{\gamma_e}{2}
\begin{bmatrix} 
1 & 0\\
0 & 1
\end{bmatrix}
\otimes
\begin{bmatrix} 
0 & B_x-iB_y\\
B_x+iB_y & 0
\end{bmatrix}.
\end{equation*}

The eigenstates of H = $H_\textrm{SO}+H_{\textrm{Z},\parallel}+H_{\textrm{Z},\perp}$ are:
\begin{equation*}
    \begin{aligned}
     &\ket{4}_{\textrm{SO}+Z,\parallel+Z,\perp} = \ket{e_+} \otimes \left[\ket{\uparrow}+\frac{\gamma_eB_+}{\gamma_eB_z+\lambda_\textrm{SO}+\sqrt{|\gamma_eB_+|^2+(\lambda_\textrm{SO}+\gamma_eB_z)^2}}\ket{\downarrow}\right];\\
     & \ket{3}_{\textrm{SO}+Z,\parallel+Z,\perp} = \ket{e_-} \otimes \left[\ket{\downarrow}+\frac{\gamma_eB_+}{-\gamma_eB_z+\lambda_\textrm{SO}+\sqrt{|\gamma_eB_+|^2+(\lambda_\textrm{SO}+\gamma_eB_z)^2}}\ket{\uparrow}\right];\\
     &\ket{2}_{\textrm{SO}+Z,\parallel+Z,\perp} = \ket{e_-} \otimes \left[\ket{\uparrow}-\frac{\gamma_eB_+}{-\gamma_eB_z+\lambda_\textrm{SO}+\sqrt{|\gamma_eB_+|^2+(\lambda_\textrm{SO}+\gamma_eB_z)^2}}\ket{\downarrow}\right]; \\
     &\ket{1}_{\textrm{SO}+Z,\parallel+Z,\perp} = \ket{e_+} \otimes \left[\ket{\downarrow}-\frac{\gamma_eB_+}{\gamma_eB_z+\lambda_\textrm{SO}+\sqrt{|\gamma_eB_+|^2+(\lambda_\textrm{SO}+\gamma_eB_z)^2}}\ket{\uparrow}\right],
    \end{aligned}
\end{equation*}
where $B_+ = B_x+iB_y$. As shown, a magnetic field perpendicular to the spin-orbit axis mixes the spin states, but does not mix the orbital states. In particular, the two qubit levels still have orthogonal orbital states. 

The Jahn-Teller effect acts only on the orbital subspace, and therefore in the spin-orbit basis can be written as:
\begin{equation*}
    H_{\textrm{JT}} = 
\begin{bmatrix} 
a & c\\
c & b
\end{bmatrix}
\otimes
\begin{bmatrix} 
1 & 0\\
0 & 1
\end{bmatrix}.
\end{equation*}

Since this term captures any general effect acting only on the orbital subspace, any strain inherent to the diamond lattice can be expressed into this term. 
The eigenstates of H = $H_\textrm{SO}+H_{Z,\parallel}+H_{\textrm{JT}}$ are:
\begin{equation*}
    \begin{aligned}
    &\ket{4}_{\textrm{SO}+Z,\parallel+\textrm{JT}} = \left[\ket{e_+}+
\frac{2c}{a-b+\lambda_\textrm{SO}+\sqrt{4c^2+(a-b+\lambda_\textrm{SO})^2}}\ket{e_-}\right] \otimes \ket{\uparrow}; \\
    & \ket{3}_{\textrm{SO}+Z,\parallel+\textrm{JT}} = \left[\ket{e_-}- 
\frac{2c}{-a+b+\lambda_\textrm{SO}-\sqrt{4c^2+(-a+b+\lambda_\textrm{SO})^2}}\ket{e_+}\right] \otimes \ket{\downarrow};\\
    & \ket{2}_{\textrm{SO}+Z,\parallel+\textrm{JT}} =  \left[\ket{e_-}+
\frac{2c}{-a+b-\lambda_\textrm{SO}-\sqrt{4c^2+(-a+b-\lambda_\textrm{SO})^2}}\ket{e_+}\right] \otimes \ket{\uparrow};\\
    &\ket{1}_{\textrm{SO}+Z,\parallel+\textrm{JT}} = \left[\ket{e_+}- 
\frac{2c}{-a+b+\lambda_\textrm{SO}+\sqrt{4c^2+(-a+b+\lambda_\textrm{SO})^2}}\ket{e_-}\right] \otimes \ket{\downarrow}.
    \end{aligned}
\end{equation*}

This shows that the $c$ component of the Jahn-Teller (and strain) effect leads to non-zero overlap in the orbital states of the two qubit levels.

\subsection{Microwave drive of spin qubit}
The Rabi rate from driving our spin qubit with a microwave drive is expected to be proportional to:
$\Omega_\textrm{MW} \propto \bra{1}H_\textrm{MW}\ket{2}$
with 
\begin{equation*}
    H_\textrm{MW} = 
\begin{bmatrix} 
1 & 0\\
0 & 1
\end{bmatrix}
\otimes
\begin{bmatrix} 
0 & 1\\
1 & 0
\end{bmatrix}
\end{equation*}
since the microwave drive preserves the orbital part of the state. Taking $\ket{1}, \ket{2}$ to be the spin qubit as defined by the spin-orbit eigenstates, $\Omega_{\textrm{MW}} = 0$ due to the the two qubit levels having orthogonal orbital states. Taking $\ket{1}, \ket{2}$ to be the spin qubit as defined by the spin-orbit, parallel Zeeman, and Jahn-Teller (including strain) eigenstates, however gives:

\begin{equation*}
    \begin{aligned}
    \Omega_\textrm{MW} &\propto \frac{2c}{-a+b-\lambda_\textrm{SO}-\sqrt{4c^2+(-a+b-\lambda_\textrm{SO})^2}}\bra{e_+}\ket{e_+}\\
    &+\frac{2c}{-a+b+\lambda_\textrm{SO}+\sqrt{4c^2+(-a+b+\lambda_\textrm{SO})^2}}\bra{e_-}\ket{e_-}\\
&=2c\left(\frac{1}{-a+b+\lambda_\textrm{SO}+\sqrt{4c^2+(-a+b+\lambda_\textrm{SO})^2}}+\frac{1}{a-b+\lambda_\textrm{SO}+\sqrt{4c^2+(a-b+\lambda_\textrm{SO})^2}}\right).
    \end{aligned}
\end{equation*}

To simplify this expression, we note that only the c term type strain contributes to the orbital mixing which enables microwave drive, and thus we consider strain with components such that $a=b$, and assume $\lambda_\textrm{SO} \gg c$, which yields:
\begin{equation*}
 \frac{4c}{\lambda_\textrm{SO}+\sqrt{4c^2+\lambda_\textrm{SO}^2}} \approx \frac{2c}{\lambda_\textrm{SO}}.
\end{equation*}

This shows that direct microwave drive of the SnV qubit is only allowed due to the Jahn-Teller (and strain) effect. However, as these energy scales (typical strains on this sample are around 10\,GHz, while the Jahn-Teller is around 65\,GHz \cite{Trusheim2020}) are much smaller than the energy scale of the spin-orbit effect (850\,GHz), driving the spin qubit via microwaves is highly inefficient. 

\begin{figure}[h]
    \centering
    \includegraphics[width=0.49\textwidth]{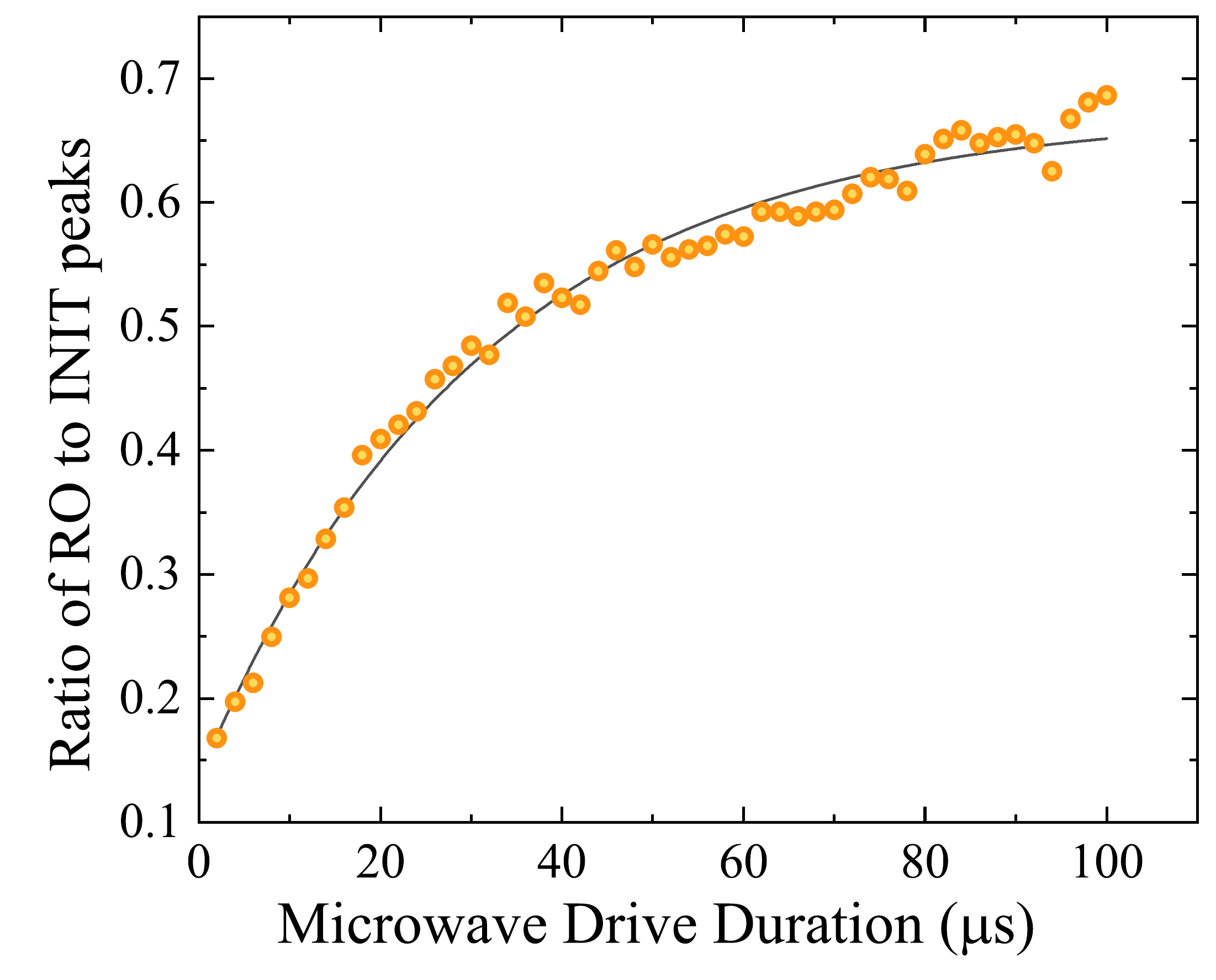}
    \caption{Microwave drive measurement. Ratio of readout pulse peak fluorescence (RO) to initialisation pulse peak fluorescence (INIT) for the emitter measured in \cite{Trusheim2020} as a function of microwave drive duration at 4\,K. SnV spin state is prepared in the $\ket{\uparrow}$ state followed by a resonant microwave drive with 10\,W of power at the cryostat entry. The solid line is a fit to $1-b(\exp[-t/a])+c$ with $a=31(2)$\,$\upmu$s, $b=0.54(1)$, $c=-0.33(1)$.}
    \label{fig:MW}
\end{figure}

\subsection{Microwave drive results}
Figure S1 shows the results of our attempts to drive an SnV with microwave drive. While this SnV is not the same SnV that is studied in the main text, it is from the same part of the device and has similar properties in terms of zero phonon line wavelengths, and ground and excited state gyromagnetic ratios. Analogously to the pulse sequence shown in Fig.\,2a, the pulse sequence used here consists of an initialize, drive, and readout pulse. The $\ket{\downarrow}$ population is calculated from the signal intensity observed in the readout pulse normalized by that observed in the initialize pulse. The drive consists of 10\,W of microwave power at a frequency resonant with the spin qubit being applied using the same microwave delivery system as in reference \cite{Trusheim2020}. Figure S1 shows the $\ket{\downarrow}$ population as a function of the microwave drive duration $T$. Fitting the recovery in $\ket{\downarrow}$ population to an exponential yields a time constant of 31(2)\,$\upmu$s, 200 times smaller than what is achieved in the main text with a lambda scheme with 200\,nW per Raman drive. Given the coherence time $T_2^*=1.3(3)\,\upmu$s found in the main text, the microwave drive would need to be over an order of magnitude faster (two orders of magnitude more microwave power) to approach the coherent control regime.

\subsection{Realizing a lambda scheme}

\begin{figure}
    \centering
    \includegraphics[width=0.49\textwidth]{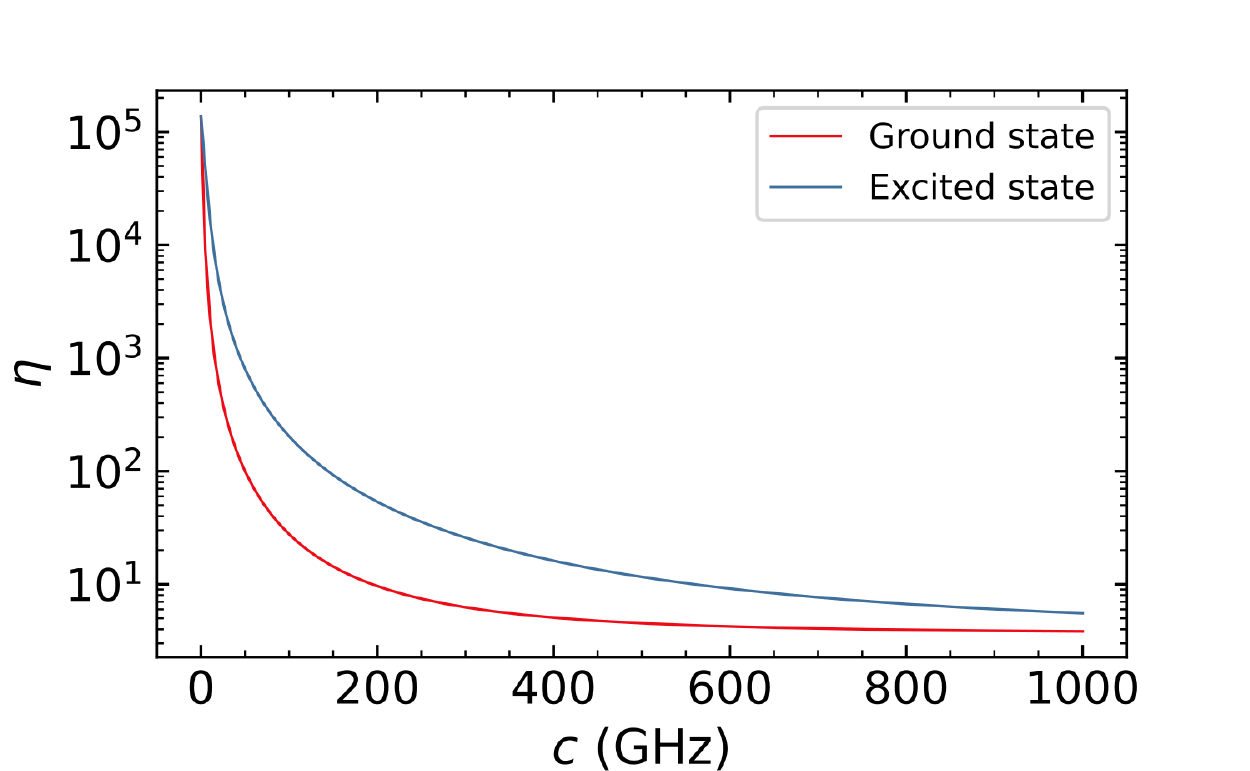}
    \caption{Branching ratio of the lambda scheme when a Jahn-Teller effect with strength $c$ is included in the ground state (red) or in the excited state (blue) for a magnetic field of 0.2\,T applied at 54.7º to the symmetry axis of the SnV.}
    \label{fig:branching_ratio}
\end{figure}

We now apply the Hamiltonian presented above to both the ground and excited states, so as to model the transitions between the two. These transitions are of particular interest as they allow for Raman transitions between the spin ground states. 

The Hamiltonian describing the excited state is composed of the same terms as that describing the ground state, presented in SI~IIIA \cite{Hepp2014}. In particular, the Zeeman effect and spin-orbit effect terms have the same matrix representations up to specific energy values. This is because in both the ground and excited states, the Zeeman effect sets a quantization axis along the direction of the applied magnetic field and the spin-orbit effect sets a quantization axis along the SnV symmetry axis \cite{Hepp2014}. The difference is that while the magnitude of the Zeeman effect is similar in both the ground and excited states, the magnitude of the spin-orbit effect is much higher in the excited state. Specifically, the spin-orbit effect has a strength of $\sim$3000\,GHz in the excited state, and largely dominates over the Zeeman splitting ($\sim$4\,GHz at $B=0.2$\,T) \cite{Trusheim2020}. Given these two terms, the quantization axis for the electronic spin in the excited state is pinned to the symmetry axis of the SnV, and thus $\ket{\textrm{A}}=\ket{e_+\downarrow}$. In the ground state, the spin-orbit contribution is weaker, and perturbs the eigenstates, which to first order become $\ket{1}\approx\ket{e_+}\otimes\big[ \ket{\downarrow}-\frac{\gamma_eB_+}{2\lambda_\textrm{SO}}\ket{\uparrow}\big] $ and $\ket{2}\approx\ket{e_-}\otimes\big[ \ket{\uparrow}-\frac{\gamma_eB_+}{2\lambda_\textrm{SO}}\ket{\downarrow}\big] $

The dipole operator enabling optical transitions driven by unpolarized light between the excited state $\ket{\textrm{A}}$ and the ground state qubit levels $\ket{1}$ and $\ket{2}$ acts as the identity in the spin basis and has components in the orbital subspace given in the ${\ket{e_+}, \ket{e_-}}$ basis by \cite{Hepp2014}
\begin{equation*}
    \hat{p}_x = 
\begin{bmatrix} 
0 & 1\\
1 & 0
\end{bmatrix},
\hat{p}_y = 
\begin{bmatrix} 
0 & -i\\
i & 0
\end{bmatrix},
\hat{p}_z = 
\begin{bmatrix} 
2 & 0\\
0 & 2
\end{bmatrix}.
\end{equation*}

Thus, from Fermi's golden rule, the expected strength of the A1 ``spin conserving" optical transition is proportional to:
\begin{equation*}
    \begin{aligned}
    |\bra{A}\hat{p}\ket{1}|^2~&=|\bra{A}\hat{p}_x\ket{1}|^2+|\bra{A}\hat{p}_y\ket{1}|^2+|\bra{A}\hat{p}_z\ket{1}|^2\\
    &=|\bra{A}\hat{p}_z\ket{1}|^2\\
    &=4,
    \end{aligned}
\end{equation*}
and that of the A2 ``spin flipping" optical transition is proportional to:
\begin{equation*}
    \begin{aligned}
    |\bra{A}\hat{p}\ket{2}|^2~&=|\bra{A}\hat{p}_x\ket{2}|^2+|\bra{A}\hat{p}_y\ket{2}|^2+|\bra{A}\hat{p}_z\ket{2}|^2\\
    &=2~\left|\bra{\downarrow}~\left( \ket{\uparrow}-\frac{\gamma_eB_+}{2\lambda_\textrm{SO}}\ket{\downarrow}\right)\right|^2\\
    &=\frac{\gamma_e^2 B_+^2}{2 \lambda_\textrm{SO}^2}.
    \end{aligned}
\end{equation*}

This demonstrates that under an off-axis magnetic field the ``spin-flipping" A1 transition becomes allowed. Nevertheless, given the large $\lambda_\text{SO}$ of SnV, this transition is still largely suppressed. Figure \ref{fig:branching_ratio} presents a numerical simulation of the branching ratio when a Jahn-Teller term with $a=b=0$ and varying $c$ is also included \cite{Trusheim2020}. It shows that the Jahn-Teller effect in the ground or excited state can explain the more balanced fraction in the Rabi rates of $\sim 100$ measured in this work. As this term can also be induced by strain, the branching ratio can vary between different emitters.

Driving these two transitions results in a lambda scheme between the two qubit states and a shared excited state which can be used to drive the spin qubit all-optically. Whereas control techniques relying on directly driving the spin qubit magnetically face the challenge of the spin qubit having near orthogonal orbital degrees of freedom, the all-optical control technique circumvents this issue by relying on the optical electric dipole moments featuring more relaxed orbital selection rules. The efficacy of this scheme depends on (1) the ratio of the decay rates for the A1 and A2 transitions, a parameter we define as $\eta$, and (2) how well light is able to couple to the SnV and drive these transitions, quantified by the saturation power of the A1 transition $p_\textrm{sat}$.

\section{Characterization of tin-vacancy lambda scheme parameters}
\label{SI:branching}
\begin{figure}
    \centering
    \includegraphics[width=0.49\textwidth]{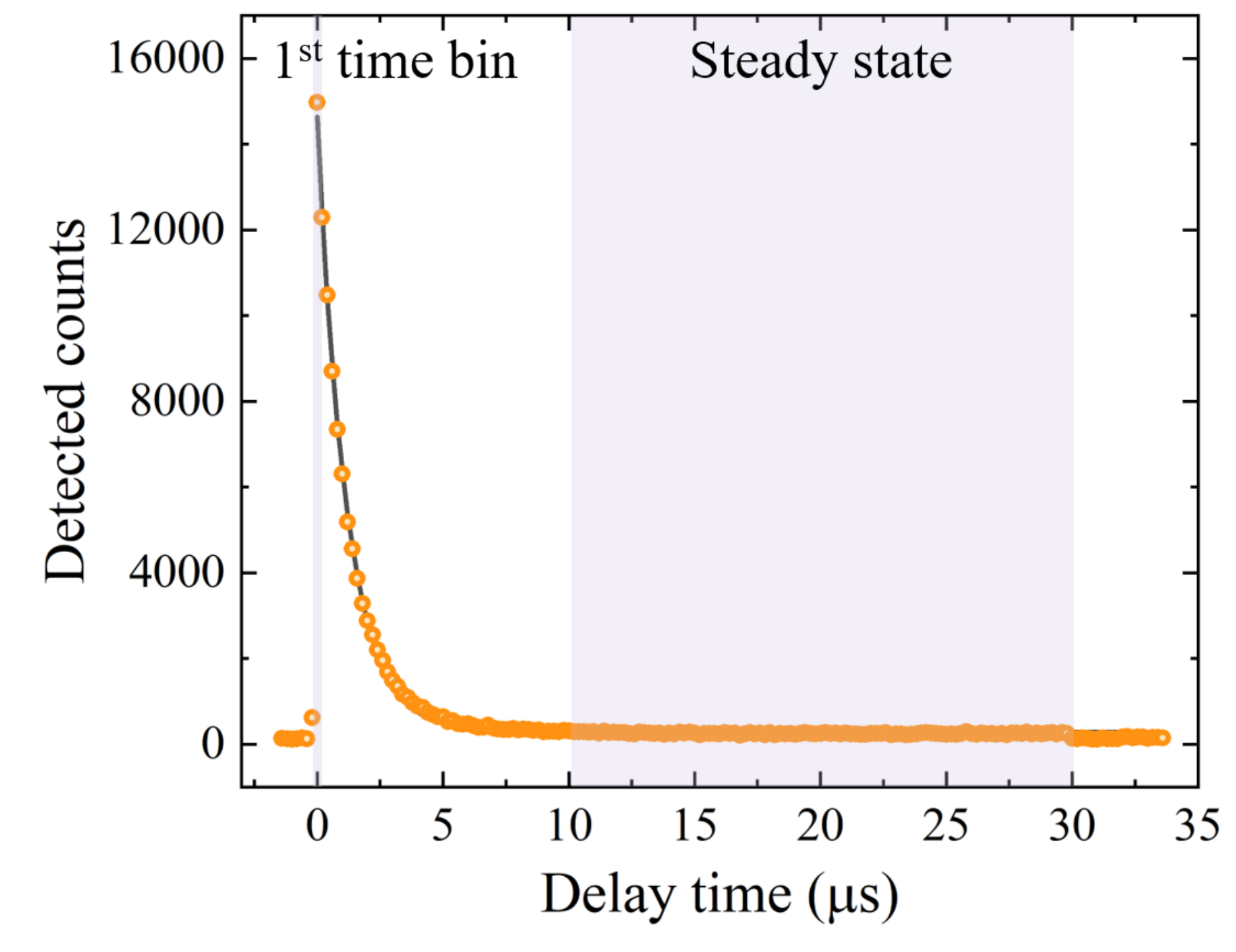}
    \caption{Fluorescence during resonant drive measurement. APD counts during a 30\,$\upmu$s long A1 pulse plotted in orange circles as a function of time. The solid curve is a fit to $a \textrm{exp}[-t/\tau_\textrm{init}]+c$ resulting in $\tau_\textrm{init} = 1.2(1)$\,$\upmu$s. The first time bin includes 14968 counts, the average of the steady state time bins is 281 counts,  and the background is 141 counts.}
    \label{fig:init}
\end{figure}

In this section, we analyze the time-resolved counts obtained during an initialize pulse and use this to obtain values for the initialization fidelity as well as $\eta$ and $p_\textrm{sat}$. 
Figure\,\ref{fig:init} shows an example of the fluorescence measured during an initialize pulse (preceded by a reset pulse to ensure $\sim$100\% population in the $\ket{\downarrow}$ state), where a resonant laser drives A1. The fluorescence signal is proportional to the population being driven by the A1 drive. The signal decreases exponentially over time as population decays from the $\ket{1\downarrow}$ state to the $\ket{2\uparrow}$ state via the spin-flipping transition A2. Taking the ratio of the fluorescence in the first time bin to that in the steady state time bins, minus background counts, we find $\epsilon$\,=\,0.9\% of the  population remaining in the $\ket{\downarrow}$ state. We thus extract an initialization fidelity $F_\textrm{init}=1-\frac{\epsilon}{2}\,=\,99.6$\%. This measured value of initialization fidelity could be limited by resonant laser leakage past our 633\,nm longpass filter and by off-resonant excitation of the B2 transition, both of which prevent the steady state counts from dropping to zero. The latter issue could be suppressed by working at higher magnetic fields or lower powers such that off resonant excitation of the B2 transition is lowered.

\begin{figure}
    \centering
    \includegraphics[width=0.49\textwidth]{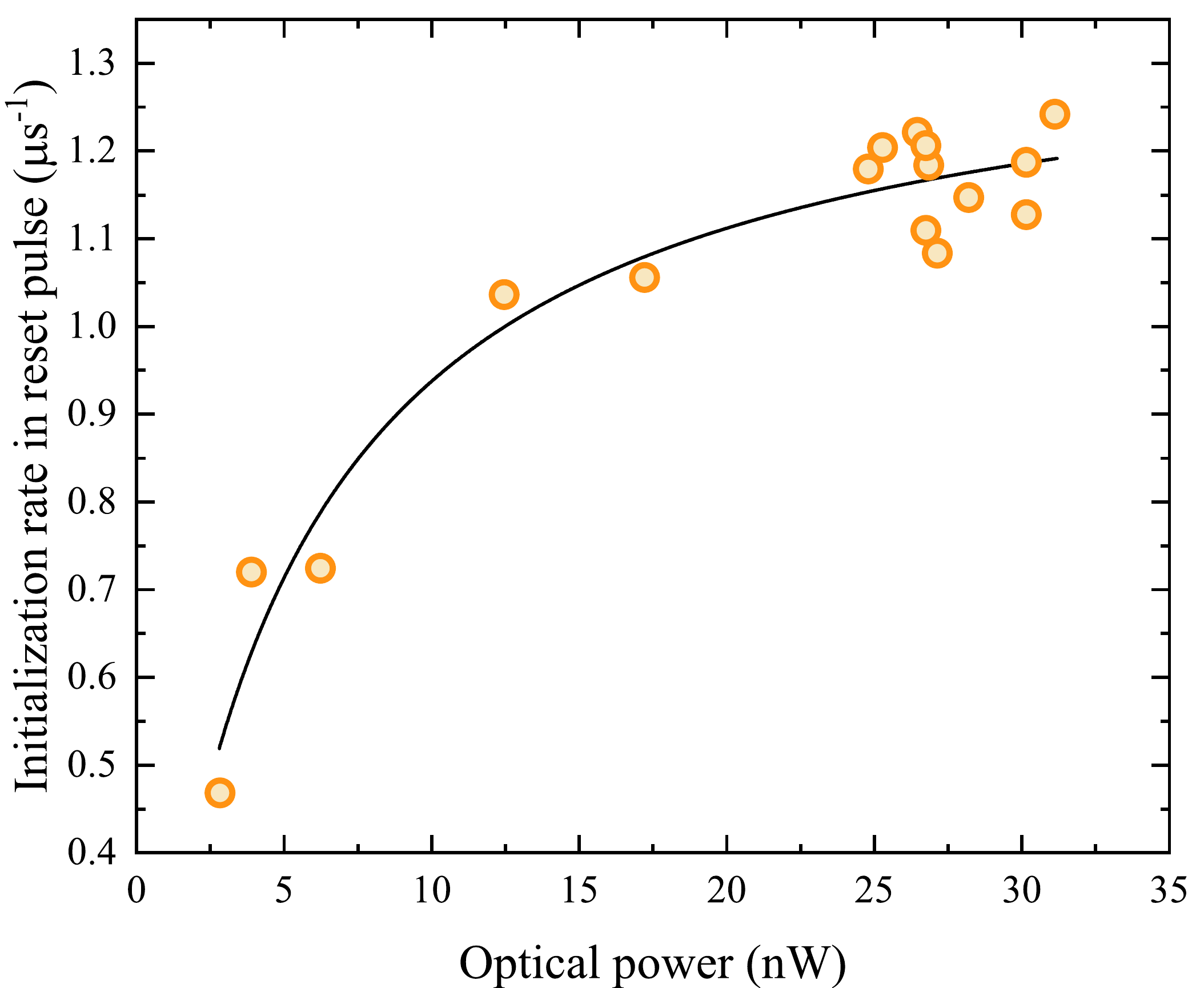}
\caption{Initialization rate for different laser powers. The initialization rates are plotted in orange circles as a function of initialization drive laser power. The solid curve is a fit to $\frac{\Gamma}{2} \frac{p/p_\textrm{sat}}{1 + p/p_\textrm{sat}} \frac{1}{\eta}$ resulting in $p_\textrm{sat}\,=\,4.6(7)$\,nW and $\eta\,=\,80(5)$.}
    \label{fig:inittimeconstants}
\end{figure}

The initialization measurement explained above is repeated at various laser powers $p$. For each $p$, the fluorescence counts are fit to an exponential decay to extract the initialization rate, defined as the inverse of the exponential time constant, presented in \fref{fig:inittimeconstants}. Fitting the initialization rates to $\frac{\Gamma}{2}  \frac{p/p_\textrm{sat}}{1 + p/p_\textrm{sat}} \frac{1}{\eta}$ \cite{Foot}, we find $p_\textrm{sat}\,=4.6(7)$\,nW and $\eta\,=\,80(5)$.

\section{Modelling all-optical coherent control}
\label{SI:RabiRate}
In this section, we first present an analytical model describing all-optical Rabi and apply it to the Rabi measurements presented in Fig.\,2a. We then develop a master equation model, and compare the results from the analytical model to those extracted by fitting the master equation model to the measured data. The consistency between the models confirms our understanding of the physics captured in the Rabi measurements. Finally, we comment on a key metric of our all-optical gates: their gate fidelity. 
\subsection{Analytical model}
We consider the lambda system presented in Fig.\,1a, consisting of two spin ground states $\ket{\downarrow}$ and $\ket{\uparrow}$ and an excited state $\ket{\textrm{E}}$ with a natural linewidth $\Gamma$ driven by two laser fields at frequencies $\omega_{1,2}$. The optical Rabi rate achieved by $\omega_1$, driving the spin-cycling optical transition A1, is given by  $\Omega_1 = \frac{1}{\sqrt{2}}\sqrt{s}\Gamma$, where $s=\frac{p}{p_\textrm{sat}}$ is the saturation parameter \cite{Foot}. The Rabi rate achieved by $\omega_2$, driving the weakly allowed spin-flipping optical transition A2, is given by $\Omega_2 = \frac{1}{\sqrt{\eta}} \Omega_1 = \frac{1}{\sqrt{2\eta}}\sqrt{s}\Gamma$. By driving $\omega_{1,2}$ simultaneously at a single photon detuning $\Delta$, the two ground states are driven with a Rabi rate of $\Omega = \frac{\Omega_1 \Omega_2}{2 \Delta} = \frac{s\Gamma^2}{\sqrt{\eta}4\Delta}$. 

Driving $\omega_{1,2}$ will also result in scattering off the excited state at a rate given by $\Gamma_{\textrm{os}} = \frac{s\Gamma^3}{8\Delta^2}$ in the limit of large $\Delta$ ($\Delta > \sqrt{s}\Gamma$). Each scattering event results in a phase given by $\omega_e/\Gamma$ being accrued, and therefore leads to decoherence. Due to this, the spin coherence time cannot exceed $T_{2,\textrm{os}} =\frac{1}{\Gamma_{\textrm{os}}}$. This phenomenon also sets an upper bound on the spin lifetime $T_{1,\textrm{os}} = \frac{\eta}{\Gamma_{\textrm{os}}}$, as one in every $\eta$ scattering events results in a spin flip.

We now apply the equations derived in our analytical model, $\Omega = \frac{s\Gamma^2}{\sqrt{\eta}4\Delta}$ and $\Gamma_{\textrm{os}} = \frac{s\Gamma^3}{8\Delta^2}$, to the case of the Rabi measurements in Fig.\,2a. In these equations, we set $\eta=80$ and $p_\textrm{sat}=4.6$\,nW as found in SI section II. The single photon detuning is set to $\Delta/2\pi=1200$\,MHz, and $p$ is set to 650\,nW, as in the measurements presented in Fig.\,2a. Using these values, the analytical model predicts $\Omega/2\pi=3.9(9)$\,MHz and $\Gamma_\textrm{os}=3.3(6)$\,$\upmu \textrm{s}^{-1}$.

\subsection{Master equation model}
We will now describe a 2-level model under a master equation formalism that can be fit to the Rabi measurements presented in Fig~2a. In this model, the von Neumann equation acquires a non-unitary term, known as the Lindbladian super-operator, which models Markovian decoherence dynamics \cite{Lindblad1976}:

$$\frac{\partial \rho}{\partial t} = \frac{-i}{\hbar} [H,\rho] + \sum_i c_i\rho c_i^\dagger -\frac{1}{2}(c_i^\dagger c_i\rho + \rho c_i^\dagger c_i),$$
where $\{c_i\}$ are the set of collapse operators and $\rho$ is the density operator. In this model, unitary evolution is driven by the Hamiltonian:

$$H=\frac{\Omega}{2}\sigma_x + \frac{\delta}{2}\sigma_z,$$
where $\sigma_i$ is the $i$'th Pauli matrix, $\Omega$ the Rabi rate and $\delta$ the two-photon detuning. The collapse operators are given by:

$$c_1=\sqrt{\frac{\gamma_{1}}{2}}\sigma_x;$$
$$c_2=\sqrt{\frac{\gamma_{2}}{2}}\sigma_z,$$
where $c_1$ describes $T_1$ depolarisation, with depolarisation rate $\gamma_{1}$ and $c_2$ describes $T_2$ pure dephasing, with pure dephasing rate $\gamma_{2}$. To account for non-Markovian inhomogenous dephasing mechanisms, leading to $T_2^*$ limited coherence times without dynamical decoupling, a phenomenological model was adopted. In this model two-photon detunings are sampled from a normal distribution with standard deviation $ \frac{1}{T_2^*}$ \cite{barry2020sensitivity, fujiwara2020real}, and averaged together; just as how slow inhomogenous dephasing manifests itself experimentally. Finally, to account for $T_1$ depolarisation amongst the 4-level electro-nuclear manifold on the time-scales probed in Fig.\,2 of the main text, a Rabi-visibility depolarisation term of the form $\frac{1}{4}(1-e^{-T\frac{\gamma_{1}}{2\pi}})$ was included in the model. In this formalism, $T$ is the Raman drive time in Fig.\,2, and this response is added to the 2-level model to yield the final model with free parameters $\Omega$, $\gamma_1$ and $\gamma_2$.

Fitting this master equation model to the Rabi measurements (as presented in Fig.\,2a), we obtain $\Omega/2\pi=3.6(1)$\,MHz and a pure dephasing rate, $\gamma_{T_2}$, of 7(4)$\,\upmu \textrm{s}^{-1}$. As the analytically derived $\Gamma_\textrm{os}$ is within the confidence interval of the extracted pure dephasing rate, we conclude that the pure dephasing rate is dominated by the optical scattering process described above. The consistency between the analytically derived expressions and the measured data confirms that the analysis of section A captures the salient physics involved in our all-optical control scheme. 

\subsection{Gate fidelities}
We will now comment on a key metric of the gates realized via our all-optical control scheme, the $\pi/2$ gate fidelity. For a $\pi/2$ gate limited by a $T_1$ process such as optical scattering, the gate fidelity can be defined as $F_{\pi/2} = 0.5(1+\textrm{exp}(-1/Q_{\pi/2})]$, where $Q_{\pi/2} = (16/9)\Omega/\Gamma_\textrm{os}$. Using the $\Omega/2\pi=3.6(1)$\,MHz and $\Gamma_\textrm{os}=7(4)$\,$\upmu$s$^{-1}$ values reported in Fig.\,2a, we calculate $F_{\pi/2}=0.92(4)$. As this value is limited by optical scattering, it could be improved by increasing $\Delta$.

Figure SI 5 shows the expected fidelity of gates performed at various $s$ and $\Delta$. To include infidelities introduced by inhomogeneous dephasing, we extend the definition presented above to $Q_{\pi/2} = (16/9)\Omega/\Gamma_\textrm{tot}$, where $\Gamma_\textrm{tot} = \textrm{max}[\Gamma_\textrm{os},\,1/T_2^*]$, where $T_2^*\,=\,1.3\,\upmu$s as found in Fig.\,3a. For fixed $s$, in the regime $\Delta < \sqrt{s\Gamma^3T_2^*/8}$ the gate fidelity is limited by optical scattering, whereas for $\Delta > \sqrt{s\Gamma^3T_2^*/8}$ the gate fidelity is limited by the quasi-static noise bath responsible for $T_2^*$.

\begin{figure}[h]
    \centering
    \includegraphics[width=0.49\textwidth]{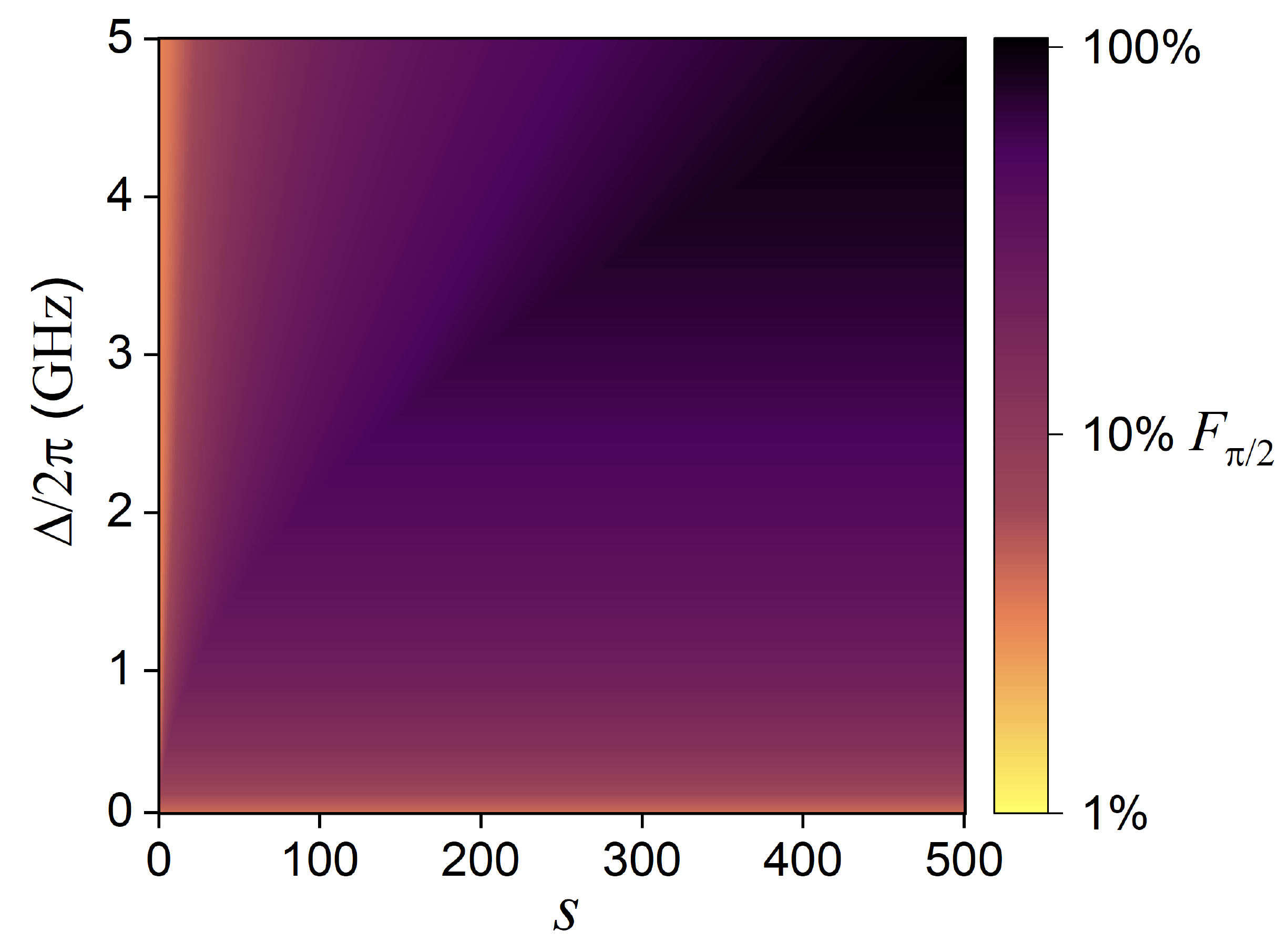}
    \caption{Rabi gate fidelity simulation. Plot of the $\pi/2$ gate fidelity $F_{\pi/2}$ as a function of $\Delta$ and $s$.}
    \label{fig:infidelity}
\end{figure}

\section{Sample fabrication}
\label{SI:sampleFab}
\begin{figure}
    \centering
    \includegraphics[width=0.35\textwidth]{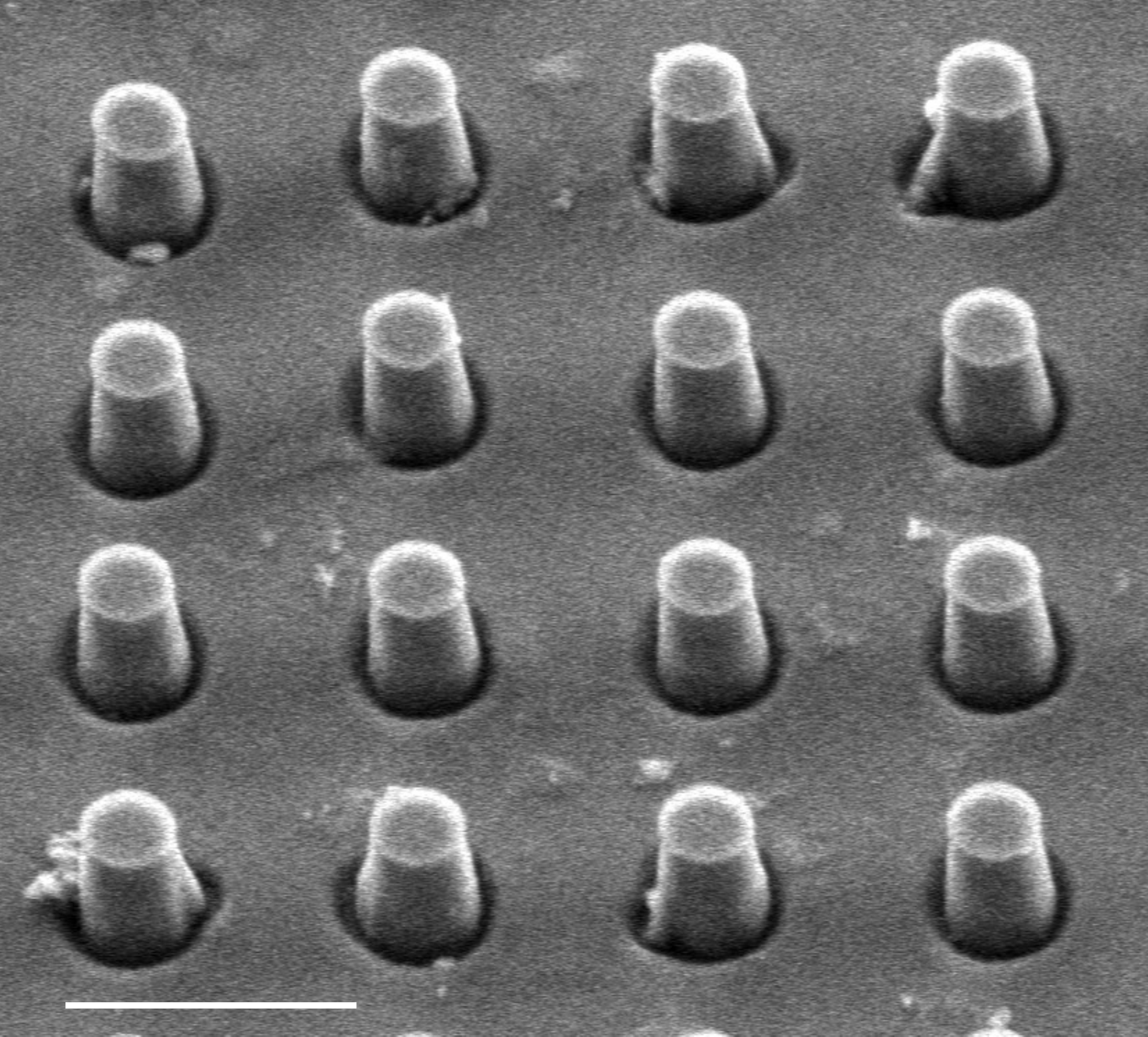}
    \caption{Scanning electron micrograph of the fabricated nanopillars. Scale bar is 1\,$\upmu$m.}
    \label{fig:pillars}
\end{figure}

The sample is the same as that used in Ref. \cite{Trusheim2020}. The diamond is an Element6 CVD-grown type IIa diamond with $<$ 5\,ppb [N],[B]. Sn+ ions were implanted with a fluence of $10^9$\,ions/cm$^{2}$ and at an energy of 350\,keV for a predicted dopant depth of 80(10)\,nm below the diamond surface as modeled by Stopping Range of Ions in Matter simulations. The sample was then annealed at $1200^\circ$ C for two hours under high vacuum ($<10^{-7}$\,mbar) and subsequently cleaned in a 1:1:1 mixture of boiling sulfuric, nitric and perchloric acid to remove residual graphite. Finally nanopillars (radii 75 to 165\,nm), shown in \fref{fig:pillars}, were fabricated to improve fluorescence collection efficiency and to isolate individual SnV centers.
\section{Equipment set-up}

\subsection{Experimental details on primary optical system}
\begin{figure*}[!h]
    \centering
    \includegraphics[width=0.5\textwidth]{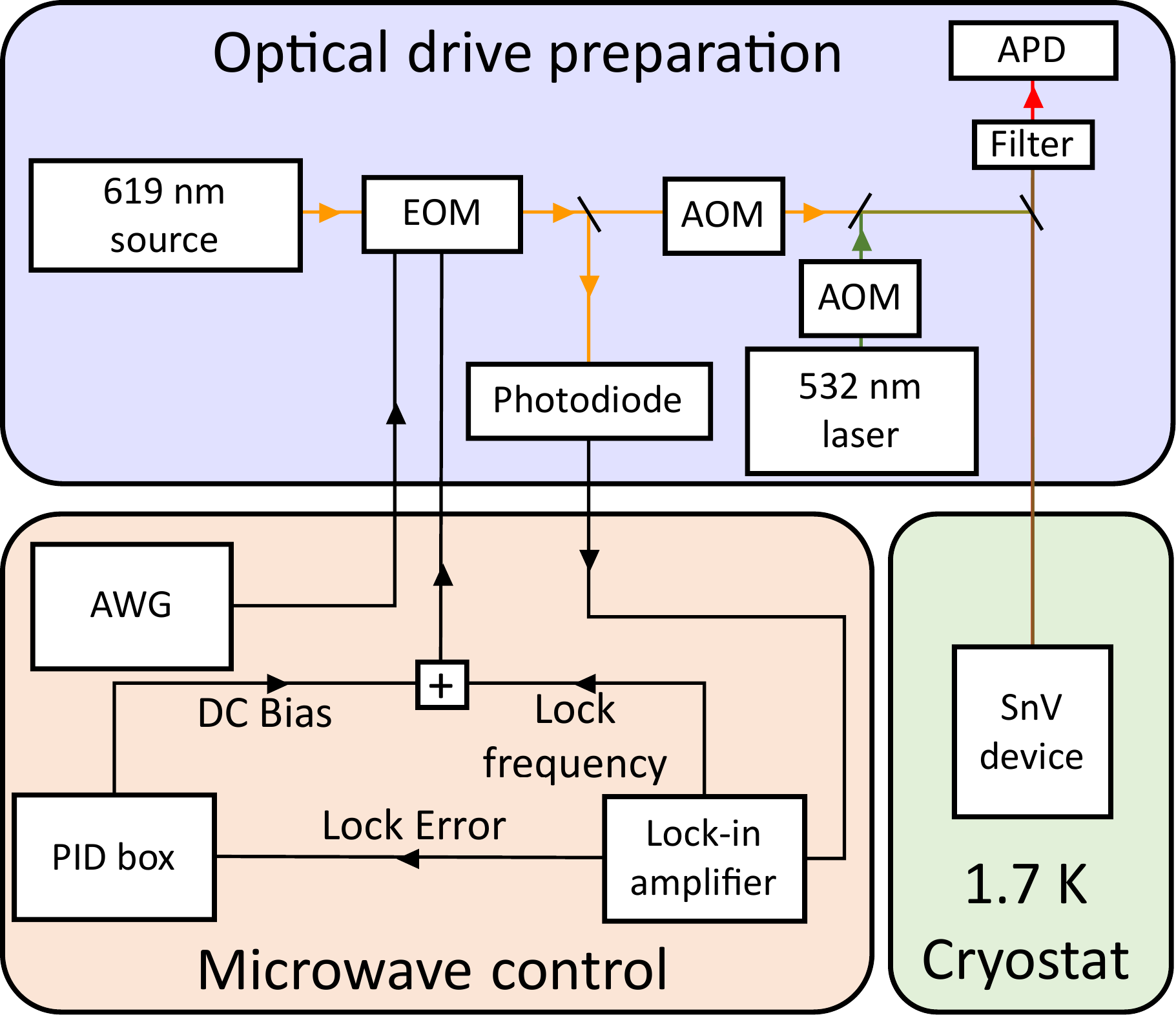}
    \caption{Schematic of experimental setup. Optical drive preparation in the blue panel: 
    619\,nm light passes through an EOM to generate sidebands. Pulse timings for the 619\,nm and 532\,nm lasers are implemented by acoustic-optic modulators (AOMs). Emission from the sample is separated from the excitation path using a 70:30 non-polarizing beamsplitter. The light is filtered by a 633\,nm long-pass filter and sent to an avalanche photodiode (APD). Microwave control in the orange panel includes an arbitrary waveform generator (AWG) as well as a lock-in amplifier and PID box to stabilize the EOM setpoint. The optical cryostat in the green box cools sample to 1.7\,K and enables the application of magnetic fields. }
    \label{fig:setup}
\end{figure*}
The optical fields used to drive the lambda system are two sidebands generated on a single laser source by an amplitude electro-optic modulator (Jenoptik AM635), and the amplitude, phase, and frequency of the sidebands are controlled by a 25 Gs$/$sec arbitrary waveform generator (Tektronix AWG70002A). Reset and initialize/readout pulses are generated by changing the AWG microwave frequency such that the EOM sidebands are resonant with the A2 and A1 transitions respectively. The EOM is locked to its interferometric minimum by modulating it with a reference signal from a lock-in amplifier (SRS SR830 DSP Lock-in Amplifier) at f = 9.964\,kHz with a feedback loop on the signal generated by a photodetector (Thorlabs PDA100A2). The error signal is sent to a PID (SRS SIM960 Analog PID Controller), whose output is applied to the EOM by a bias-tee.

The data shown in the measurements presented in this work were taken in a closed-cycle cryostat (attoDRY 2100) with a base temperature of 1.7\,K at the sample and in which the temperature can be tuned with a resistive heater located under the sample mount. Superconducting coils around the sample space allow the application of a vertical magnetic field from 0 to 9\,T and a horizontal magnetic field from 0 to 1\,T. Unless explicitly stated otherwise, all measurements were conducted at T=1.7\,K and B=0.2\,T with the magnetic field orientation 54.7\,$^{\circ}$ rotated from the SnV symmetry axis. The optical part of the set-up consists of a confocal microscope mounted on top of the cryostat and a microscope objective with numerical aperture 0.82 inside the cryostat. The sample is moved with respect to the objective utilizing piezoelectric stages (ANPx101/LT and ANPz101/LT) on top of which the sample is mounted. Resonant excitation around 619\,nm is performed by a second harmonic generation stage (ADVR RSH-T0619-P13FSAL0) consisting of a frequency doubler crystal pumped by a 1238\,nm diode laser (Sacher Lasertechnik Lynx TEC 150). The frequency is continuously stabilized through feedback from a wavemeter (High Finesse WSU). The charge environment of the SnV- is reset with microsecond pulses at 532\,nm (M-squared Equinox). Optical pulses are generated with an acousto-optic modulator (Gooch and Housego 3080-15) controlled by a delay generator (Stanford Research Instruments DG645). For resonant excitation measurements, a long-pass filter at 630\,nm (Semrock BLP01-633R-25) is used to separate the fluorescence from the phonon-sideband from the laser light. The fluorescence is then sent to a single photon counting module (SPCM-AQRH-TR), which generates TTL pulses sent to a time-to-digital converter (Swabian Timetagger20) triggered by an arbitrary waveform generator (Tektronix AWG70002A). Photon counts during ``initialize" and ``readout" pulses are histogrammed in the time-tagger to measure the $\ket{\downarrow}$ population as described in the main text. 

\subsection{Experimental details on second optical system}

The data in \fref{fig:CPT} is measured in a BlueFors He dilution refrigerator at 3.2\,K. The sample consists of an SnV implanted in a diamond microchiplet with several waveguides. A lensed fiber (OZ Optics TSMJ-3U-1550-9/125-0.25-7-2.5-14-2) is used to collect PL, which is filtered to measure the phonon sideband. The sample is placed in a static magnetic field $\sim$ 0.1\,T produced by a permanent magnet. Coherent population trapping (CPT) is measured using the same pulse sequence described in the main text but with rest on A2 instead of A1.

\subsection{Measurement conditions for main-text data}
The coherent control used throughout the main text requires that we measure the spin Rabi rate to determine the drive duration required to produce $\pi/2$ and $\pi$ rotations. In Table \ref{tab:parameter}, we present a summary of the single-photon detuning $\Delta$ and power-per-sideband $p$ used in each figure.

\begingroup
\setlength{\tabcolsep}{8pt} 
\begin{table}[h]
\begin{tabular}{@{}rccccccc@{}}
\toprule
\textbf{Main text figure} & \textbf{1b} & \textbf{1c} & \textbf{2a} & \textbf{2b} & \textbf{3} & \textbf{4 (Hahn echo)} & \textbf{4 (CPMG-2)} \\ \midrule
$\Delta / 2\pi$\,(MHz)     & 0           & 600         & 1200        & 300         & 300        & 600                    & 800                 \\
$p$\,(nW)                  & 40          & 40          & 650         & 260         & 260        & 200                    & 200                 \\ \bottomrule
\end{tabular}
\caption{Table of measurement parameters, single-photon detuning $\Delta$ and power $p$ per sideband  with uncertainty taken to be 10\%, for data presented in the main text. }
    \label{tab:parameter}
\end{table}
\endgroup

\section{Coherent population trapping model and simulations}
\label{SI:CPT}
\begin{figure}[!h]
    \centering
    \includegraphics[width=0.49\textwidth]{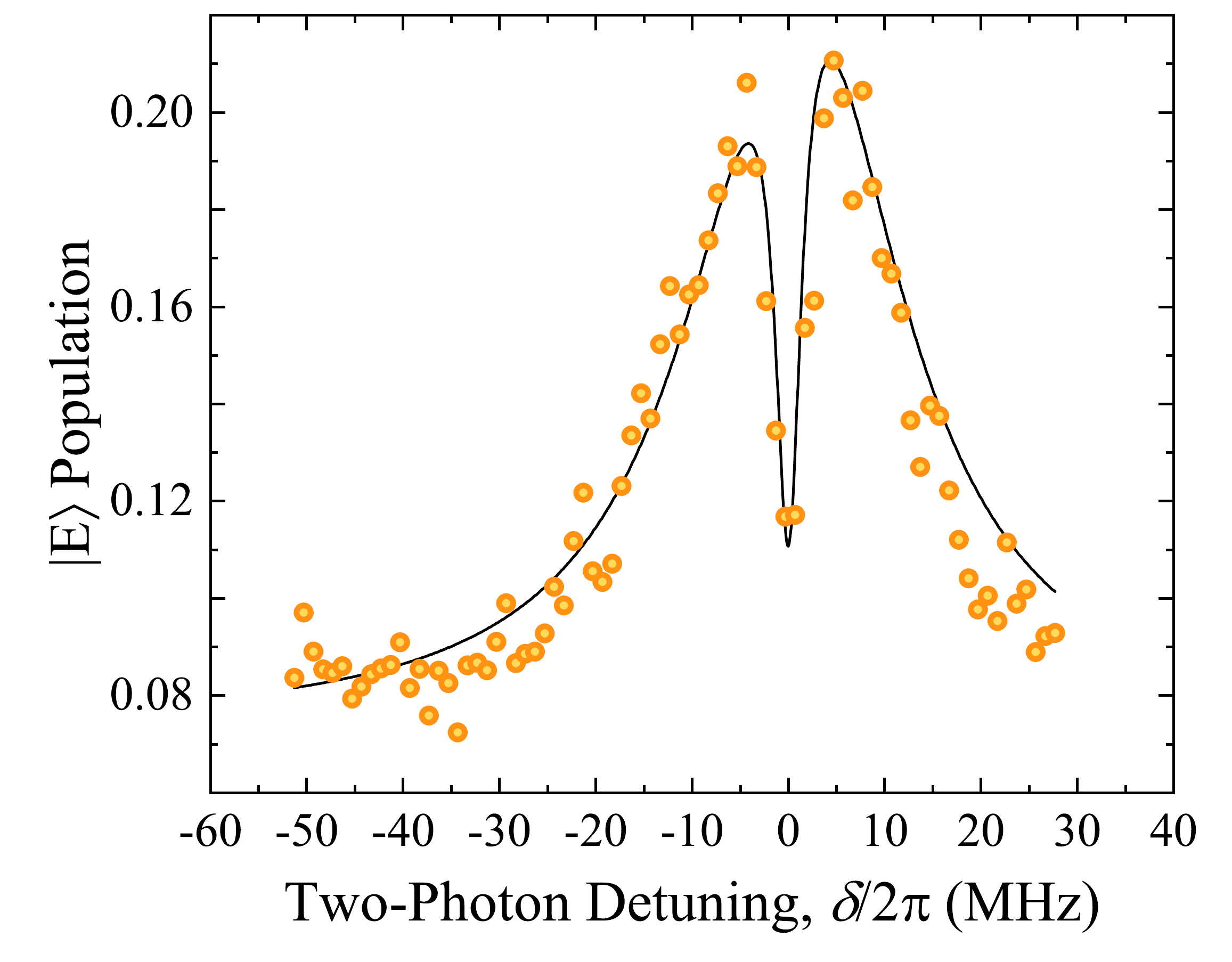}
    \caption{Coherent population trapping measured on an SnV without a hyperfine splitting and $\sim 600$\,nW per sideband. The solid curve is a fit to the aforementioned CPT model with the following free-parameters: $\gamma_{\downarrow,\uparrow}=0.2(1)$\,MHz, $\gamma_{\downarrow \downarrow,\uparrow \uparrow}=2.72(7)$\,MHz, $\Omega=8.0(6)$\,MHz and $\Delta=0.9(2)$\,MHz.}
    \label{fig:CPT}
\end{figure}
To model the CPT outlined in Fig.\,1b of the main text, we use a three level quantum system, where $\ket{\downarrow}$ and $\ket{\uparrow}$ are the ground states and $\ket{\textrm{E}}$ is the excited state. In the rotating-frame of the $\ket{\downarrow} \leftrightarrow \ket{\textrm{E}}$ the Hamiltonian is given by \cite{RevModPhys.77.633}:

$$\hbar\begin{pmatrix}
-\Delta & 0 & \frac{\Omega_1}{2}\\
0 & -\delta & \frac{\Omega_2}{2}\\
\frac{\Omega_1}{2} & \frac{\Omega_2}{2} & 0
\end{pmatrix},$$
where $\Delta$ is the single-photon detuning, $\delta$ is the two-photon detuning and $\Omega_1$ ($\Omega_2$) is the Rabi rate between $\ket{\downarrow}$ ($\ket{\uparrow}$) and the excited state $\ket{\textrm{E}}$. Given the cyclicity of SnV, the two Rabi rates are related by $\frac{\Omega_1}{\sqrt{f}}=\frac{\Omega_2}{\sqrt{1-f}}$, where $f$ is the acyclicity of the lambda scheme and is related to the branching ratio, $\eta$, by $\frac{f}{(1-f)}=\frac{1}{\eta}$.

This Hamiltonian drives unitary evolution of the system's state-vector between its eigenstates, as determined by the von Neumann equation. At two-photon resonance, these eigenstates coalesce into:
\begin{equation*}
\begin{aligned}
&\ket{B_+} = \sin\theta\sin\phi\ket{\downarrow}+\cos\phi\ket{\textrm{E}}+\cos\theta\sin\phi\ket{\uparrow}\\
&\ket{B_{-}} = \sin\theta\cos\phi\ket{\downarrow} - \sin\phi\ket{\textrm{E}}+\cos\theta\cos\phi\ket{\uparrow}\\
&\ket{D} = \cos\theta\ket{\downarrow} - \sin\theta\ket{\uparrow},
\end{aligned}
\end{equation*}
where $\tan \theta=\frac{\Omega_1}{\Omega_2}$ and $\tan 2\phi =\frac{\sqrt{\Omega_1^2 + \Omega_2^2}}{\delta}$. Crucially, the new ground state $\ket{D}$ is orthogonal to the excited state $\ket{\textrm{E}}$ and is therefore dark. Accordingly, scattering from the lambda scheme pumps population into the dark state, as seen by an absence of counts at two-photon resonance.

However, non-unitary dissipative dynamics interfere with the coherence of the lambda scheme and enable scattering from the excited state, even at two-photon resonance. As the generator of Markovian dissipative dynamics, the Lindblad master-equation is suited to model such decoherence dynamics \cite{Lindblad1976}:

$$\frac{\partial \rho}{\partial t} = \frac{-i}{\hbar} [H,\rho] + \sum_i c_i\rho c_i^\dagger - \frac{1}{2}(c_i^\dagger c_i\rho + \rho c_i^\dagger c_i),$$
where $\{c_i\}$ are the set of collapse operators and $\rho$ is the density operator. Accordingly, the pure-dephasing collapse-operator is given by:

$$\sqrt{\frac{\gamma_{\downarrow\downarrow,\uparrow\uparrow}}{2}}(\ket{\downarrow}\bra{\downarrow}+\ket{\uparrow}\bra{\uparrow}),$$
where $\gamma_{\downarrow\downarrow,\uparrow\uparrow}$ is the inhomogeneous dephasing rate, modelled here as a pure-dephasing rate.
Scattering rates from the excited state into $\ket{\downarrow}$ and $\ket{\uparrow}$ are given by
$$\sqrt{\frac{\gamma_{\downarrow,\textrm{E}}}{2}}\ket{\downarrow}\bra{\textrm{E}},$$
$$\sqrt{\frac{\gamma_{\uparrow,\textrm{E}}}{2}}\ket{\uparrow}\bra{\textrm{E}},$$
respectively. The cyclicity of the SnV enforces the constraint $\frac{\gamma_{\downarrow,\textrm{E}}}{f}=\frac{\gamma_{\uparrow,\textrm{E}}}{1-f}$, which is equal to the excited-state scattering rate. Finally, the external bath coupled to the SnV centre induces relaxation parameterised by:
$$\sqrt{\frac{\gamma_{\downarrow,\uparrow}}{2}}(\ket{\downarrow}\bra{\uparrow}+\ket{\uparrow}\bra{\downarrow}),$$
where $\gamma_{\downarrow,\uparrow}$ is the $T_1$ dephasing rate.
Fitting this model recreates the data presented in Fig.\,1b and implies that the hyperfine-interaction induces a 43.6(8)\,MHz shift in the two-photon detuning between the electro-nuclear spin manifold. The remaining fit parameters are shown to be:
$$\frac{\Omega_1}{\sqrt{f}}=\frac{\Omega_2}{\sqrt{1-f}}=24.7(1.0) \text{\,MHz}$$
$$\Delta=-2.72(58) \text{\,MHz}$$
and an excited state decay rate $\Gamma/2 \pi=32(8)$\,MHz.

In \fref{fig:CPT}, we furthermore provide a measure of CPT on an SnV in a device measured at 3.2\,K in a BlueFors He4-He3 dilution refrigerator. The data in \fref{fig:CPT} was measured using the same sequence in Fig.\,1b of the main text, but with reset on A2 in stead of on A1.

\section{Intensity-correlation measurement under resonant excitation}

\label{SI:g2}
\begin{figure}[!h]
    \centering
    \includegraphics[width=0.49\textwidth]{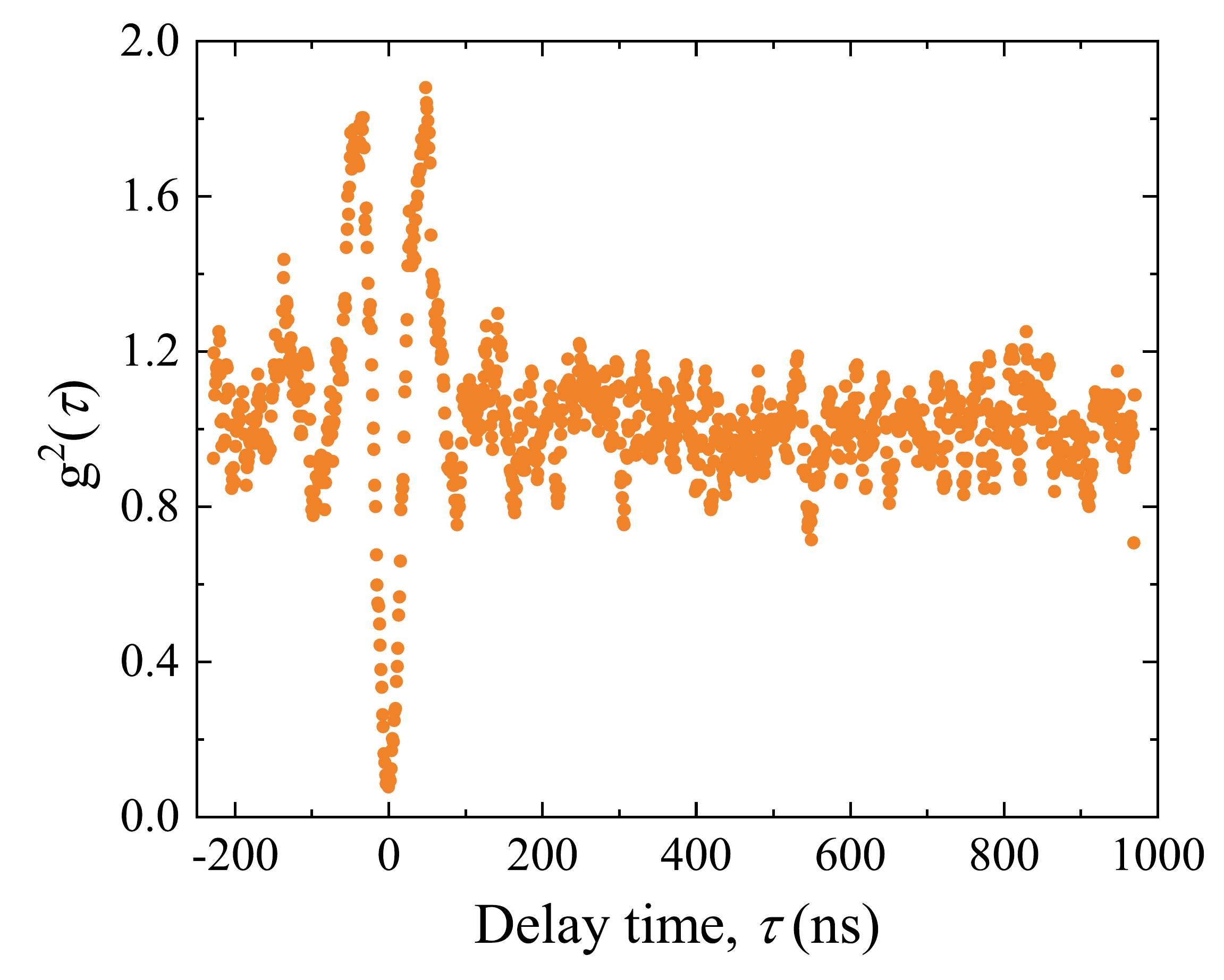}
    \caption{Intensity auto-correlation g$^2$($\tau$) measurement. $g^2(\tau)$ measured as a function of $\tau$, plotted as orange circles and connected by a solid line. This measurement was taken at 0 magnetic field such that all four optical transitions (A1, A2, B1, B2) are simultaneously driven with a resonant laser. At $\tau =0$, the measured autocorrelation value reaches a minimum value of $g^2(0)$ = $0.08 (1)$.} 
    \label{fig:g2}
\end{figure}
Figure\,\ref{fig:g2} presents a g$^2$($\tau$) auto-correlation measurement on the SnV featured in the main text. The measured $g^2(0) = 0.08(1)$ unambiguously confirming that we measure a single SnV. We collect emission into the phonon sideband (as in main text) and excite the emitter with continuous-wave resonant excitation at zero magnetic field.

\section{Ramsey signal analysis}
\label{SI:Ramsey}

\subsection{Serrodyne frequency $\omega_\textrm{S}$}
The serrodyne frequency causes the phase of the second $\pi$/2 pulse, $\phi$, to be modulated periodically as a function of the delay duration $\tau$, $\phi=\omega_\textrm{S}\tau$. In the absence of two-photon detuning and AC Stark effect ($\delta=\Delta_\textrm{AC}=0$), this would rotate the axis of the second $\pi$/2 pulse at a rate $\omega_\textrm{S}$, and therefore introduce oscillations in the $\ket{\downarrow}$ population with frequency $\omega_{\textrm{Ramsey}} = \omega_\textrm{S}$.
\subsection{Two-photon detuning $\delta$}
When the two photon detuning is nonzero, the rotating frame of the drive, set by $\omega_R$, is no longer the same as the rotating frame set by $\omega_e$. Therefore, in the rotating frame set by $\omega_e$, the drive axis precesses at a rate $\delta=\omega_R-\omega_e$. Including the effects of both the serrodyne frequency and detuning, the angle between the axis of the state and the axis of rotation is $\phi+\delta(\tau+2T_{\pi/2})=\tau(\omega_\textrm{S}+\delta)+2 \delta T_{\pi/2}$, where $2 T_{\pi/2}$ is the cumulative duration of the two $\pi/2$ pulses. This results in $\omega_{\textrm{Ramsey}} = \omega_\textrm{S} + \delta$. The $2\delta T_{\pi/2}$ term captures the precession due to the two-photon detuning during the $\pi$/2 pulses, and explains the phase shift between different horizontal line cuts observed in Fig.\,3b.

\subsection{Differential AC Stark effect $\Delta_{\textrm{AC}}$}
The differential AC Stark effect arises due to the difference in AC Stark shifts experienced by the two spin-cycling transitions (A1 and B2). While the drive pulse is on, the spin qubit splitting is increased by $\Delta_\textrm{AC}$ and is what we measure as $\omega_e$. During the delay time, when the spin qubit is no longer subject to the drive pulse causing the AC Stark shifts, the spin qubit splitting is $\omega_e-\Delta_\textrm{AC}$. This means that in the $\omega_e$ rotating frame, the spin qubit will precess at a rate $-\Delta_\textrm{AC}$ during the delay time. The accumulated angle between the axis of the state and the axis of rotation is then $\phi+\delta(\tau+2T_{\pi/2})+\Delta_\textrm{AC}\tau=\tau(\omega_\textrm{S}+\delta+\Delta_\textrm{AC})+2\delta T_{\pi/2}$. This results in $\omega_{\textrm{Ramsey}} = \omega_\textrm{S} + \delta + \Delta_\textrm{AC}$.

The magnitude of $\Delta_\textrm{AC}$ can be calculated by applying the all-optical drive model presented in SI V. Using $\Omega/2\pi=1.4$\,MHz, $\eta= 80(5)$, and single-photon detuning for the spin-conserving transition $\Delta_1/2\pi = 300(5)$\,MHz, we find an optical Rabi rate $\Omega_1/2\pi=87(4)$\,MHz. The splitting in the dressed state is $\sqrt{\Omega_1^2 + \Delta_1^2}$, and the  $\ket{\downarrow}$ ground state shifts relative to the undriven case by the AC Stark shift  $\Delta_{\textrm{AC,1}} = \frac{\sqrt{\Omega_1^2 + \Delta_1^2}-\Delta_1}{2}\approx  \Omega_1^2/(4\Delta_1)$.  Then $\Delta_{\textrm{AC,1}}/2\pi = 6.3(6)$\,MHz. The same optical drive is experienced by the $\ket{\uparrow}$ spin-cycling transition but at a detuning $\Delta_2 = \Delta_1 + (\omega_{\uparrow\uparrow}- \omega_{\downarrow\downarrow})$, where $(\omega_{\uparrow\uparrow}- \omega_{\downarrow\downarrow})$ is the difference in energy between the spin-conserving transitions, 610(5)\,MHz. In this case, the shift in the $\ket{\uparrow}$ ground state is $\Delta_{\textrm{AC,2}}/2\pi \approx  \Omega_1^2/(4\Delta_2) = 2.1(2)$\,MHz, where the spin-conserving optical Rabi rates are assumed to be equal for $\ket{\downarrow}$ and $\ket{\uparrow}$ and $\Delta_2=910(10)$\,MHz. The differential AC Stark shift is then $\Delta_{\textrm{AC}} =\Delta_{\textrm{AC,1}} - \Delta_{\textrm{AC,2}}=4.2(8)$\,MHz, which we find is in reasonable agreement with the measured value for $\Delta_{\textrm{AC}} = 3.3(5)$\,MHz. We have neglected contributions from AC Stark effects due to spin-flipping transitions for which the optical Rabi rate is $\sqrt{\eta}$ times weaker, resulting in a $\sim1$\% correction to our estimate for $\Delta_\textrm{AC}$. We also neglect the AC Stark shifts due to $\omega_2$ detuned from the spin-conserving peaks by $\sim4$\,GHz, which would be a $\sim10$\% correction to our estimated value. 

\subsection{Ramsey simulation}
\begin{figure}[!h]
    \centering
    \includegraphics[width=0.99\textwidth]{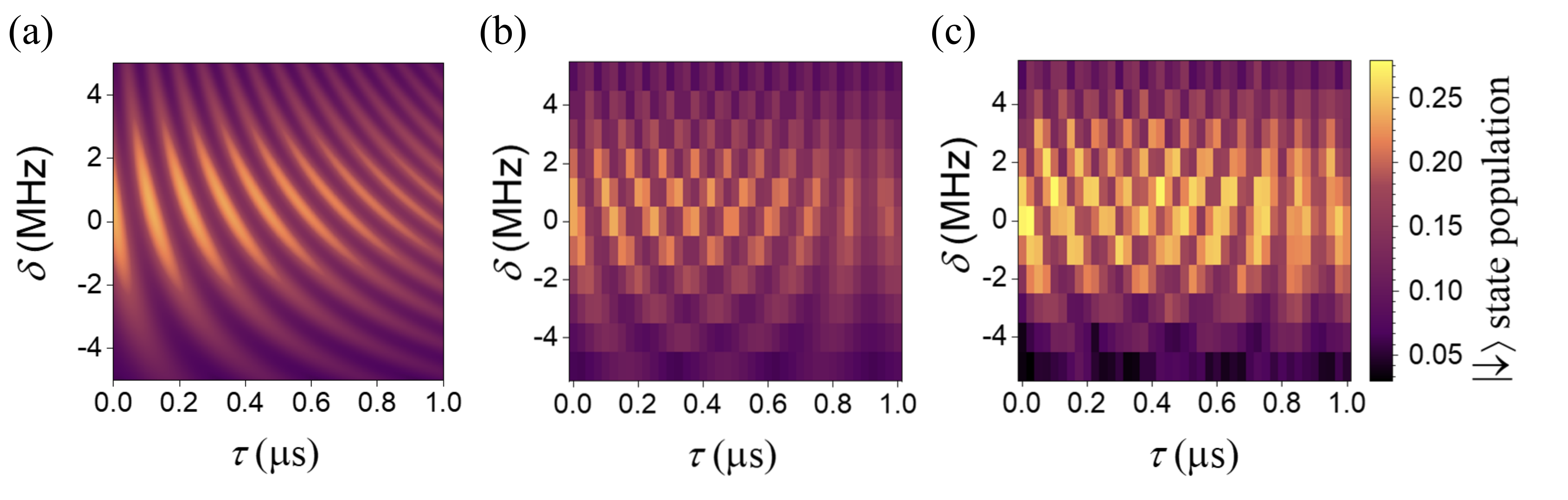}
    \caption{Ramsey fit and comparison to data. (a) Fine-step (1\,ns and 10\,kHz) simulation for $\ket{\downarrow}$ state population as a function of $\delta$ and $\tau$ obtained from the model in \eref{eqt:Ramsey}. (b) Coarse step simulation. Same as (a) but with the simulation step size for $\delta$ = 1\,MHz and the simulation step size for $\tau$ = 25\,ns, the same step sizes used in the measurement of Fig.\,3b. (c) Reproduction of Fig.\,3b data for comparison.}
    \label{fig:ramsey1MHZ}
\end{figure}
Figure\,\ref{fig:ramsey1MHZ}a plots the predicted $\ket{\downarrow}$ population obtained from the model
\begin{equation}\frac{c_0}{1+(\delta/c_1)^2}
\textrm{exp}[-(\tau/T_2^*)^2]\Big( \textrm{cos}\big( \omega_\textrm{Ramsey}\tau + 2\delta T_{\pi/2}\big)+1\Big)+\frac{c_2}{1+(\delta/c_1)^2}
\label{eqt:Ramsey}
\end{equation}
where $\omega_{\textrm{Ramsey}} = \omega_\textrm{S} + \delta + \Delta_\textrm{AC}$, $T_{\pi/2}$ is fixed to the experimentally measured $\pi$/2 gate time, $T_2^*$ is fixed to the measured inhomogeneous dephasing time found in Fig.\,3a, and $\Delta_\textrm{AC},\, c_0$,\, $c_1$, \,$c_2$ are fitting parameters. Figure\,\ref{fig:ramsey1MHZ}b evaluates the same expression at the time steps taken experimentally in Fig.\,3b (reproduced as \fref{fig:ramsey1MHZ}c), showing that certain visual artifacts in Fig.\,3b (i.e. the lack of $\ket{\downarrow}$ population at $\tau$=900\,ns for all measured $\delta$) are simply a result of the aliasing effect from a finite sampling of the time and detuning axes. 

\section{Spin lifetime $T_1$ measurement}
\begin{figure}[!h]
    \centering
    \includegraphics[width=0.5\textwidth]{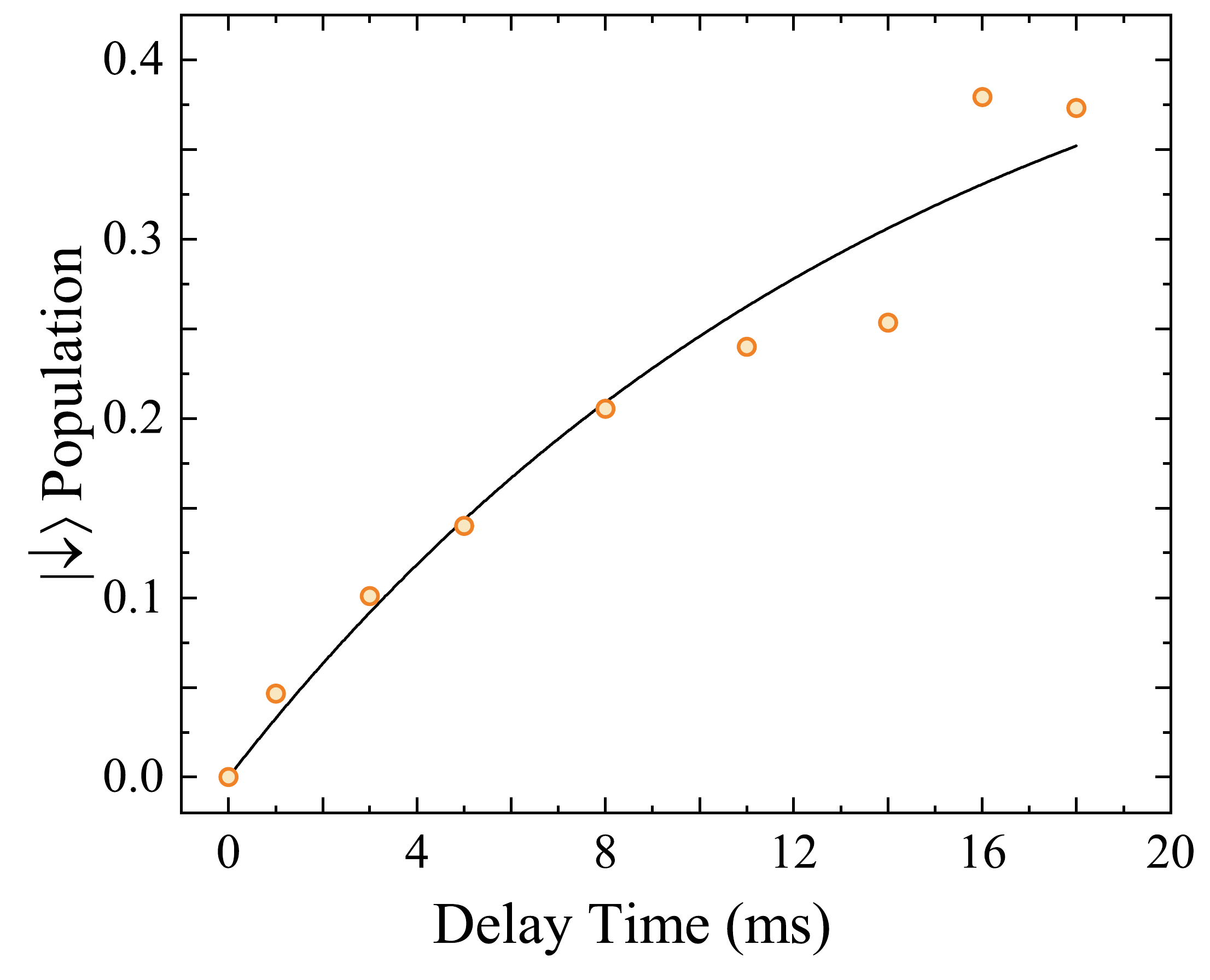}
    \caption{$T_1$ measurement under laser leakage. The pulse sequence consists of reset, initialize, delay, and readout pulses, as described in the main text. Orange circles represent $\ket{\downarrow}$ state population recovery as a function of delay time $\tau$. The recovery is fit to $0.5*(1-e^{-\tau/T_1})$ (solid line), from which we extract $T_1$ = 15(1) ms.}
    \label{fig:T1}
\end{figure}
In this section, we present spin lifetime $T_1$ measurements taken with 1.0(1)\,nW of laser leakage at 1.2\,GHz detuned from the A1 optical transition. These measurements consist of a reset, initialize, delay, and readout pulses, as described in the main text. Figure\,S10 presents the recovery of $\ket{\downarrow}$ population as a function of the delay time. We extract $T_1$=15(1)\,ms from this data. As the expected limit on $T_1$ from interorbital phonons at 1.7\,K is $T_{1,\textrm{phonons}}\,\gg1$\,s \cite{Trusheim2020}, we conclude the measured $T_1$ is not limited by phonons. In contrast, using the equations derived in SI section III we calculate that the optical scattering due to the laser leakage would limit $T_1 \le$ \,16(2)\,ms and $T_2\,\le\,$0.20(3)\,ms. As these values are with error of the $T_1$ measurement presented in this section and the $T_2$ measurement presented in Fig.\,4, we conclude that these measurements were limited by optical scattering.

\bibliography{SnVControlRefs_v1.bib}